\documentclass[fleqn,a4paper,useAMS,usenatbib]{mnras} 

\usepackage{txfonts}
\newcommand{\text}{\mathrm}  

\usepackage{ucs}
\usepackage[utf8x]{inputenc}   

\usepackage[T1]{fontenc}
\usepackage{ae,aecompl}

\usepackage{graphicx}	

\usepackage{ifthen}            
\usepackage[normalem]{ulem}    

\newcommand{\hrsim}{hMS}               
\newcommand{\lcdm}{$\Lambda$CDM}       

\newcommand{\vv}[1]{\mathbfit{#1}}      

\newcommand{\dd}{\text{d}}              
\newcommand{\dotprod}{\mathbf{\cdot}}

\newcommand{\DM}{\Delta\mu}            
\newcommand{\DMpr}{\Delta\mu_\text{pr}} 
\newcommand{\Mform}{M_\text{form}}
\renewcommand{\mp}{ m_\text{p} }
\newcommand{\Np}{N_\text{p}}
\newcommand{\Npin}{N_\text{p,inner}}
\newcommand{\Nplim}{N_\text{p,lim}}
\newcommand{\Qlim}{Q_\text{lim}}
\newcommand{\thetain}{\theta_\text{inner}}
\newcommand{\jsc}{\tilde{\jmath}}
\newcommand{\jscin}{\tilde{\jmath}_\text{inner}}
\newcommand{\jsclim}{\jsc_\text{lim}}
\newcommand{\Mh}{M_\text{h}}
\newcommand{\Rvir}{R_\text{vir}}
\newcommand{\Mvir}{M_\text{vir}}
\newcommand{\Rhalfm}{R_{1/2}}
\newcommand{\tlb}{t_\text{lb}}  
\newcommand{\fM}{f_\text{M}}
\newcommand{\Nprog}{N_\text{prog}}

\newcommand{\OmegaM}{\Omega_\text{M}}
\newcommand{\OmegaL}{\Omega_\Lambda}
\newcommand{\OmegaMzero}{\Omega_{\text{M}0}}
\newcommand{\OmegaLzero}{\Omega_{\Lambda0}}
\newcommand{\Omegabzero}{\Omega_{\text{b}0}}

\newcommand{\rhoc}{\rho_\text{crit}}   
\newcommand{\rhoczero}{ \rho_{\text{crit,}0} }
\newcommand{\Deltac}{\Delta_\text{c}}  

\newcommand{\rhos}{\rho_\text{s}}      
\newcommand{\rs}{r_\text{s}}           

\newcommand{\Msol}{\text{M}_{\sun}}      
\newcommand{\munit}{\,h^{-1}\Msol}   
\newcommand{\Mpc}{\text{Mpc}}               
\newcommand{\kpc}{\text{kpc}}               
\newcommand{\kms}{\text{km}\,\text{s}^{-1}}        
\newcommand{\Hunit}{\,\kms\,\Mpc^{-1}} 
\newcommand{\lunit}{ \, h^{-1} \Mpc }      
\newcommand{\klunit}{\, h^{-1} \kpc }      
\newcommand{\Gyr}{\, \text{Gyr} }           

\newcommand{\figw}{80 mm}     

\newcommand{\figwthree}{58mm} 
\newcommand{\figwtwo}{65 mm}  

\graphicspath{{./figs/}}
\DeclareGraphicsExtensions{.eps} 

\defcitealias{paperI}{Paper I}

\newcommand{\annotate}{false}  

\newcommand{\drafting}{newver} 

\newcommand{\pbscrapsty}[1]{\sout{#1}}
\newcommand{\pbnewsty}[1]{#1}

\ifthenelse{\equal{\annotate}{true}}{
  \newcommand{\pbtodo}[1]{ {\small[\textbf{#1}]} }
}{
  \newcommand{\pbtodo}[1]{}
}

\ifthenelse{\equal{\drafting}{oldver}}{      
  \newcommand{\pbscrap}[1]{#1}                  
  \newcommand{\pbnew}[1]{}                      
} {\ifthenelse{\equal{\drafting}{newver}}{   
    \newcommand{\pbscrap}[1]{}                  
    \newcommand{\pbnew}[1]{#1}                  
  } {
    \newcommand{\pbscrap}[1]{\pbscrapsty{#1}}   
    \newcommand{\pbnew}[1]{\pbnewsty{#1}}       
  }
}

\title[Spin Flips -- II]{Spin Flips -- II. Evolution of dark matter halo
  spin orientation, and its correlation with major mergers}
\author[P. E. Bett \& C. S. Frenk]{Philip E. Bett$^{1}$\thanks{E-mail: p.e.bett@physics.org}\thanks{Current address: Met Office Hadley Centre, FitzRoy Road, Exeter, EX1 3PB.} and Carlos S. Frenk$^{2}$ \\
  $^{1}$Argelander-Institut f\"ur Astronomie, Universit\"at Bonn, 
  Auf dem H\"ugel 71, D-53121 Bonn, Germany\\
  $^{2}$Institute for Computational Cosmology, University of Durham,
  South Road, Durham, DH1 3LE, UK
}

\date{Accepted XXX. Received YYY; in original form ZZZ}

\pubyear{2016}

\begin{document}
\label{firstpage}
\pagerange{\pageref{firstpage}--\pageref{lastpage}}
\maketitle

\begin{abstract}  
  We expand our previous study on the relationship between changes in
  the orientation of the angular momentum vector of dark matter haloes
  (``spin flips'') and changes in their mass \citep{paperI}, to cover
  the full range of halo masses in a simulation cube of length
  $100\lunit$.  Since strong disturbances to a halo (such as might be
  indicated by a large change in the spin direction) are likely also to
  disturb the galaxy evolving within, spin flips
  could be a mechanism for galaxy morphological transformation without
  involving major mergers.  We find that $35\%$ of
  haloes have, at some point in their lifetimes, had a spin flip of at
  least $45\degr$ that does not coincide with a major merger.
  Over $75\%$ of large spin flips coincide with non-major mergers;
  only a quarter coincide with major mergers.  We find a similar
  picture for changes to the inner-halo spin orientation, although
  here there is an increased likelihood of a flip occurring.  Changes
  in halo angular momentum orientation, and other such measures of
  halo perturbation, are therefore very important quantities to consider, in
  addition to halo mergers, when modelling the formation and evolution
  of galaxies and confronting such models with observations.
\end{abstract}


\section{Introduction}


One of the key quantities in the evolution of cosmic strucures and the
formation of galaxies is angular momentum.  The acquisition and early
growth of angular momentum by dynamically relaxed, overdense clumps of
matter (`haloes') can be studied using linear tidal torque theory
(\citealt{1951pca..conf..195H, 1969ApJ...155..393P,
  1970Ap......6..320D, 1970Afz.....6..581D, 1984ApJ...286...38W,
  1996MNRAS.282..436C, 1996MNRAS.282..455C}; see also
\citealt{2002MNRAS.332..325P}, \citealt{2009IJMPD..18..173S} and \citealt{2015MNRAS.452.3369C}), but
this begins to break down as structure growth becomes non-linear
\citep{1984ApJ...286...38W}.  Subsequent growth then has to be studied
using $N$-body simulations.  While research in this field
dates back many decades, the continual increase in computing power
means that recent simulations have been able to determine with great
accuracy the distribution and evolution of the halo angular momentum
amplitudes (e.g. \citealt{2001ApJ...555..240B, 2005ApJ...634...51A,
  2006ApJ...646..815S, 2007MNRAS.381...41H,2007MNRAS.375..489H,
  bett07,bett10, Maccio2007,Maccio2008, 2008ApJ...678..621K,
  2011MNRAS.411..584M}). 

In contrast, the orientation of halo angular momentum is less well
studied.  Research on this topic tends to focus on the angular
momentum direction with respect to other quantities, 
such as the shape of the halo \citep[e.g.][]{1992ApJ...399..405W,
  2005ApJ...627..647B, 2006MNRAS.367.1781A, 2006ApJ...646..815S,
  2007MNRAS.377...50H, bett07,bett10},
or the orientation of galaxies \citep[e.g.][]{ 2002ApJ...576...21V,
  2003MNRAS.346..177V, 2003ApJ...597...35C, 2006PhRvD..74l3522G,
  2009MNRAS.400...43C, 2009ApJ...702.1250R, bett10,
  2010ApJ...709.1321A, 2010MNRAS.405..274H, 2011MNRAS.415.2607D},
or larger-scale filaments and voids \citep[e.g.][]{2005ApJ...627..647B,
  2007MNRAS.381...41H, 2007MNRAS.375..489H, 2007MNRAS.375..184B,
  2008MNRAS.389.1127P, 2008MNRAS.385..867C, 2010MNRAS.405..274H,
  2012MNRAS.421L.137L}.
 \pbnew{In particular, \cite{2012MNRAS.427.3320C} found that low-mass haloes, which have grown through smooth accretion, tend to have their spin vectors aligned parallel to their nearest filament. In contrast, spins in higher mass haloes, which have experienced major mergers, tend to be perpendicular to their filaments.}\footnote{\pbnew{\cite{2012MNRAS.427.3320C, 2015MNRAS.452.3369C} refer to the transition between parallel and perpendicular alignment of spin and filament, as a halo grows, as a `spin flip', determined statistically over a large halo population.  Note that in our paper we use the term to refer to \emph{any} sudden changes in spin direction in  the lifetimes of \emph{individual} haloes.}} \pbnew{\cite{2014MNRAS.444.1453D} showed that this is also true for galaxies.   Experiencing more mergers increases the likelihood of perpendicular alignment, whereas a lack of mergers allows a galaxy spin to drift back towards parallel alignment with its filament  \citep{2014MNRAS.445L..46W}.}


The evolution of the \emph{Lagrangian} mass comprising $z=0$ haloes
has been studied by \cite{2000MNRAS.311..762S} and
\cite{2002MNRAS.332..325P}, who showed that
the spin direction changes due to non-linear evolution, with both the
average deviation from the initial direction, and the scatter in that
angle, increasing with time.  


Part of the reason for the importance of angular momentum is the
strong link it provides between halo and galaxy evolution.  In the
standard cosmological model, the matter content of the Universe is
dominated by a cold, collisionless component, cold dark matter (CDM).
In this paradigm, structures grow hierarchically, through mergers of
ever-larger objects.  Galaxies then form and evolve within these
haloes \citep{whiterees1978, 1991ApJ...379...52W}.  The more complex
physical processes available to the baryons as they cycle between gas
and stars result in galaxy evolution not being strictly hierarchical
\citep[e.g.][]{bowergalform2006}.  In models of galaxy formation, the
gas is usually assumed to have initially the same angular momentum as
the halo, which is then conserved as the gas cools and collapses to
form a disc.  Thus, the size of the galactic disc is directly related
to the dark matter halo's angular momentum \citep{1980MNRAS.193..189F,
  1998MNRAS.295..319M, Zavala_Frenk08}.  This idea is widely
implemented in semi-analytic models of galaxy formation
(\citealt{1991ApJ...379...52W}; see also the reviews of
\citealt{2006RPPh...69.3101B} and \citealt{2010PhR...495...33B}).  It
is important to note that this involves only the magnitude of the halo
angular momentum, rather than the full vector quantity.  It is the
vector that is conserved, and standard semi-analytic models at present
make no reference to the angular momentum direction.

Morphological changes in galaxies can be brought about through tidal
forces \citep{1972ApJ...178..623T}, and indeed a galactic disc can be
disrupted completely if the gravitational potential varies strongly
enough over a short timescale.  Galaxy formation models thus assume
that a sufficiently large galaxy merger event will destroy a disc,
randomising the stellar orbits and forming a spheroid
\citep[e.g.][]{1977egsp.conf..401T, 1988ApJ...331..699B,
  1992ApJ...393..484B,1996ApJ...471..115B, 1992ApJ...400..460H,
  1993ApJ...409..548H}.  This has been shown also to occur in
numerical simulations of individual mergers
\citep[e.g.][]{2003ApJ...597..893N, 2005A&A...437...69B,
  2006ApJ...650..791C,2008MNRAS.384..386C}.  However, the details of
the merger process, and the properties of the resulting galaxies,
depend strongly on the gas richness of the participants
\citep[e.g.][]{2008ApJ...683..597S, 2009ApJ...702..307S,
  2009ApJ...691.1168H, 2009MNRAS.397..802H, 2010ApJ...715..202H,
  2010MNRAS.403.1009M}, and on the details of the star formation and
feedback processes triggered by the merger
\citep[e.g.][]{2005MNRAS.363.1299O,
  Zavala_Frenk08,2009MNRAS.396..696S}.

In \cite{paperI} (hereafter \citetalias{paperI}), the authors put
forward the idea that sudden, large changes in the \emph{direction} of
the halo angular momentum vector (hereafter halo \emph{spin}
direction, for brevity) are indicative of a significant disturbance to
the halo, and that, although such changes are usually assumed to only
accompany halo mergers, they can also occur without the large mass
gain implied by a merger.  This would mean that galaxies could be
distrupted by processes that are not captured in the galaxy and halo
merger trees used in most current modelling, and that sudden changes to the
halo angular momentum direction could be a useful proxy to detect such
events. 

Such events have been seen in $N$-body and hydrodynamical simulations.
\cite{2005MNRAS.363.1299O} found that their simulated disc galaxy
flipped its orientation \citep{2010AIPC.1240..403B}, with subsequent
misaligned gas accretion resulting in a transformation into a bulge,
with another disc forming later.  \cite{2009MNRAS.396..696S} also
found that misalignment of a stellar disc with accreting cold gas can
result in bulge formation, sometimes destroying the disc, and
sometimes with a new disc forming later.  The idealized experiments of
\cite{2013MNRAS.428.1055A} showed clearly the strong and complex
impact of gas/halo misalignment on the evolution of the disc and bulge
components of a halo's central galaxy.  
\cite{2009ApJ...702.1250R} analysed haloes in simulations both with
and without baryons, and found that halo spin orientations can change
much more drastically than the angular momentum magnitude, and that
such large orientation changes are not restricted to major mergers.
\pbnew{\cite{2015arXiv150205053W}  also found that it is not just major mergers that can destroy disks, although minor mergers are statistically less likely to do so; the gas content in the galaxies also has an important role.} \pbscrap{This was confirmed again by Welker et al. (2014), who focused on
the impact of spin flips and mergers on galaxy orientation within to
the cosmic web.}

Following these studies, \cite{2014MNRAS.443.2801P} used the
distribution of spin flips from haloes in the Millennium~II simulation
\citep{2009MNRAS.398.1150B} to incorporate their impact stochastically
in a semi-analytic galaxy formation model.  In their model, flips in
galaxy discs act to reduce the disc spin in proportion to the cosine
of the orientation change, i.e. larger flips reduce the spin of the
disc more.  Their model also allows disc instabilities, which cause
bursts of star formation, to be triggered by spin flips, rather than
by halo mergers explicitly.  The authors demonstrate the positive
impact that these changes to the model have on the distributions of
galaxy properties at different redshifts, such as galaxy luminosity
functions, morphological distributions and star formation rates.

In the present paper, we study the relationship between changes in
spin orientation and halo merger history, expanding on
\citetalias{paperI}.  We include haloes of all masses in the $N$-body
simulation we use, and pay special attention to those haloes that
do not survive to $z=0$; we also  compare our results against the
assumptions of mass and angular momentum conservation used in simple
halo models.  The results emphasize the need for models of galaxy
formation to include more information than just the halo mass accretion
history when determining galaxy properties, and in particular the
transfer of material from disc to spheroidal structures.

In Section~\ref{s:analy} we describe the simulation used, the
identification and selection of haloes and their merger trees, and the
properties on which we focus in our analysis.  Section~\ref{s:results}
describes our results, both in terms of the distribution of events
over all haloes, and during each halo's lifetime, including the impact
on inner halo spin directions.   We summarise our conclusions in
Section~\ref{s:concs}.


\section{Simulation data and analysis}\label{s:analy}
We use the same simulation and methods for analysis as in
\citetalias{paperI}.  While we describe the important points here, we
refer the reader to that paper for further details.

\subsection{The hMS simulation, haloes and merger trees}
We use a cosmological dark matter $N$-body simulation that has been
referred to as \hrsim{},\footnote{Other studies using the \hrsim{}
  simulation include \cite{neto07}, \cite{2008MNRAS.387..536G},
  \cite{2009MNRAS.398.1150B}, \cite{2009MNRAS.399..550L}, and
  \cite{bett10}.} since it was made using the same \textsc{L-Gadget-2}
code and \lcdm{} cosmological parameters as the Millennium Simulation
\citep[MS,][]{mill2005}, but with a smaller cube size ($100 \lunit$)
and higher resolution ($900^3$ particles of mass $9.518\times
10^7\munit$, and softening $2.4\klunit$).  The \hrsim{} assumes the
same cosmological parameters as the MS: writing cosmological density
parameters as $\Omega_i(z) = \rho_i(z) / \rhoc(z)$, in terms of the
mass density\footnote{The equivalent mass-density of the cosmological
  constant $\Lambda$ can be written as $\rho_\Lambda = \Lambda c^2/
  (8\upi G)$.} of component $i$ and the critical density $\rhoc(z) =
3H(z)^2 / (8\upi G)$, where the Hubble parameter is $H(z)$, for the
present-day cosmological constant, total mass, and baryonic mass, the
\hrsim{} uses values of $\OmegaLzero \equiv \OmegaL(z=0) = 0.75$,
$\OmegaMzero=0.25$, and $\Omegabzero = 0.045$ respectively.  The
present-day value of the Hubble parameter is parameterised in the
standard way as $H_0=100h \Hunit$, where $h=0.73$.  The spectral index
is $n = 1.0$ and the linear-theory mass variance in $8\lunit$ spheres
at $z=0$ is given by $\sigma_8 = 0.9$.

We use a halo definition based on linking and separating subhaloes
from their associated friends-of-friends (FoF) particle groups according to
information in the halo merger trees.  The algorithms for both the
haloes and merger trees were originally described in
\cite{harker2006}, and designed for use with the implementation of the
\textsc{Galform} semi-analytic galaxy formation model in the
Millennium Simulation \citep{bowergalform2006}\footnote{In particular,
  they correspond to the \texttt{DHalo} tables in the Millennium
  Simulation database \citep{milldb}}.  Particle groups are first
identified through the Friends of Friends (FoF) algorithm
\citep{defw85}, with a linking length parameter of $b=0.2$.
Self-bound substructures -- the main bulk of the halo itself, plus any
subhaloes -- are then identified using the \textsc{Subfind} code
\citep{subfind2001}.  Each halo-candidate particle group then consists
of a main self-bound structure, zero or more self-bound subhaloes,
plus additional particles that are spatially linked through FoF.

The bound subtructures can be tracked between the simulation
snapshots, identifying progenitors and descendents (see
e.g. \citetalias{paperI}, \citealt{harker2006} or
\citealt{2014MNRAS.440.2115J} for details).  Using this additional
evolution information the halo catalogue is refined, by separating off
subhaloes that are spatially but not dynamically linked to the halo.
For example, subhaloes that are just passing through the outskirts of
a larger halo are separated.  Two haloes joined by a thin bridge of
particles (as commonly occurs with FoF) would also be split apart.
\cite{bett07} compared the spins, shapes, clustering and visual
appearance of these `merger-tree haloes' against both simple FoF
groups and haloes defined using a simple spherical overdensity
criterion, and showed that this merger-tree method offers a great
improvement in terms of identifying the genuine physical structure of
a halo.

\begin{figure} 
    \includegraphics[width=\figw]{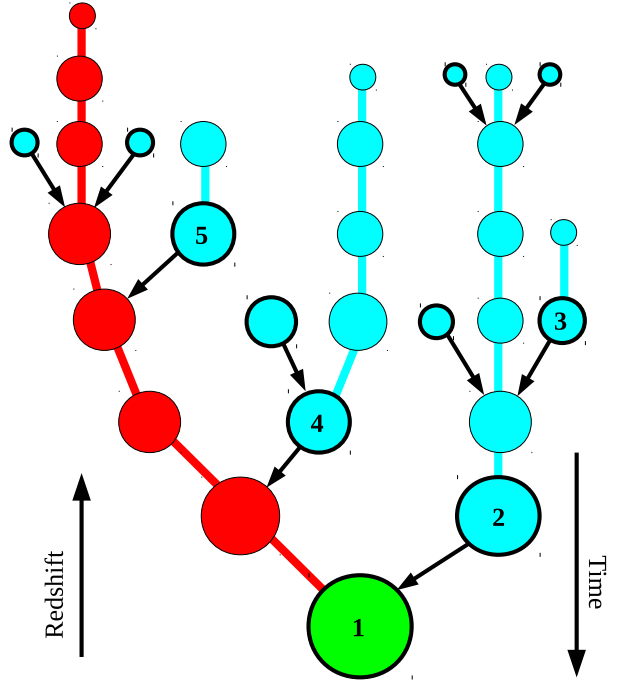}
    \caption{Schematic of a single merger tree.  Each circle
      represents a halo, with each row representing a single
      simulation output time.  The green ``root'' halo is that
      identified at $z=0$.  The evolutionary track of the most-massive
      progenitors of the root halo is coloured in red.  Tracks of
      haloes that merge into a larger halo before $z=0$ are coloured
      cyan.  End-points of tracks have a heavy black outline, with
      their final merger into a larger neighbour marked with a black
      arrow.  Six of these 11 tracks only exist at a single timestep,
      leaving five (marked with numbers) with useful halo-evolution
      data. 
}
  \label{f:mergtree}
\end{figure}

The result of the merger tree and halo identification algorithms is
a set of  haloes (groups of self-bound structures) identified at
each snapshot, with at most one descendent and zero or more
progenitors. Each halo identified at $z=0$ is the root of its own
tree, which branches into many progenitor haloes at earlier timesteps.
In this paper, we wish to study how properties of individual haloes
evolve.  We therefore identify the evolutionary ``track'' of a halo,
by finding the most massive of its immediate progenitors at each
timestep.  We give an illustration of halo tracks in a merger tree in
Fig.~\ref{f:mergtree}.  The track of the root halo is marked in red:
each red halo is the most massive progenitor of its descendent halo.
The halo population at a given snapshot is made up of many other
haloes than the root halo however: tracks of other haloes exist, which
do not survive until $z=0$. Instead, they merge into a more massive
halo at some point.  The endpoint of each track in
Fig.~\ref{f:mergtree} is highlighted in heavy black lines.  Also note
that some tracks have no evolution information, since they only exist
at one timestep before merging into a larger halo.

It is important to note that neither the halo definition nor the merger tree
algorithm are by any means unique.  The halo merger trees used  \pbnew{in the semi-analytic galaxy formation models of  \cite{mill2005} and \cite{2007MNRAS.375....2D}, which also use the Millennium Simulation, }
\pbscrap{for
those semi-analytic galaxy formation models in the Millennium
Simulation developed at the Max Planck Institute for Astrophysics
(e.g. Springel et al. 2005 and De Lucia \& Blaizot 2007)}
 construct
both the haloes and merger trees differently, although based on
similar principles.  Other methods that use splitting/stitching
algorithms similar to the one used here include those by
\cite{2002ApJ...568...52W}, \cite{2008MNRAS.386..577F,
  2009MNRAS.394.1825F}, \cite{2009ApJ...701.2002G},
\cite{2010MNRAS.403..984N}; see also \cite{2006ApJ...647..763M}.
There have recently been various detailed \pbnew{comparison} studies of halo definition and merger tree algorithms, including \cite{2009A&A...506..647T,
  2013MNRAS.435.1618K, 2013MNRAS.436..150S, 2014MNRAS.441.3488A}.  \pbnew{There is therefore significant scope for similar studies to ours to produce quantitatively different results.} 

\subsection{Halo property catalogues}
As in \citetalias{paperI}, various properties of the haloes are
computed at each snapshot in time, in the centre-of-momentum frame of
each halo, and in physical rather than comoving coordinates. The halo
centre is identified with the location of the gravitational potential
minimum of its most massive structure, as found by
\textsc{Subfind}. Properties are computed using the halo particles
only (rather than the set of particles within a certain radius), and
include the halo mass $M$, kinetic and potential
energies\footnote{Following \cite{bett07}, the potential energy $U$ is
  computed using a random sample of $1000$ particles if the halo has
  more than that.} $T$ and $U$, and angular momentum vector $\vv{J}$.
An approximate ``virial'' radius $\Rvir$, is found by growing a sphere
around the halo centre until the enclosed density from halo particles
 drops below $\Deltac(z)\, \rhoc(z)$.  The threshold
overdensity with respect to critical, $\Deltac(z)$, is given by 
the spherical collapse model \citep{ECF96}, using the fitting formula
of \cite{BN98}:
\begin{equation}
  \label{e:virdens}
  \Deltac(z) 
  = 18\upi^2  + 82\left(\OmegaM(z)-1\right) - 39\left(\OmegaM(z)-1\right)^2
\end{equation}
In the case of the flat \lcdm{} universe here, we can write
$\OmegaM(z) = \OmegaMzero a^{-3} / \chi(z)$ and $\rhoc(z) =
\rhoczero\, \chi(z)$, where the expansion factor $a = (1+z)^{-1}$ and
we define $\chi(z) = \OmegaMzero a^{-3} + \OmegaLzero$ for
convenience.  Note that we only use $\Rvir$ as a convenient spatial scale
for the haloes, rather than as a halo boundary.  We also define an
inner-halo region at $r_\text{inner} = 0.25\Rvir$ (\pbscrap{similar to} \pbnew{following the orientation resolution tests of}
\citealt{bett10}).  Using the mass within this radius, we also compute
the inner angular momentum vector $\vv{J}_\text{inner}$.

\subsection{Halo selection}\label{s:sel}
We need to select haloes at each timestep from which reliable
measurements of angular momentum can be made.  We follow the
\textit{Three Rs} of selecting haloes from simulations, requiring them
to be well-\emph{r}esolved, approximately \emph{r}elaxed, and
\emph{r}obust against effects caused by using discrete particles.
These are realised through the following selection criteria,
respectively:
\begin{eqnarray*}
  \Np            & \geq & \Nplim             \\
  \left|Q\right| & \leq & \Qlim,              \\
  \log_{10} \jsc & \geq & \log_{10} \jsclim,
\end{eqnarray*}
where $\Np$ is the number of particles comprising the halo, the energy
ratio $Q=2T/U +1$ approximates the virial ratio, and $\jsc = j/\sqrt{G
  \Mh \Rvir}$ represents a scaling of the halo angular momentum
magnitude with respect to that of a particle orbiting under gravity at
the virial radius ($j = J/M$ is the specific angular
momentum).\footnote{Note that $\jsc$ is identical to the alternative
  spin parameter $\lambda'$ introduced by \cite{2001ApJ...555..240B},
  modulo a factor of $\sqrt{2}$.}  Following \pbnew{resolution tests in} the studies of
\cite{bett07,bett10}, we choose limiting values of $\Nplim = 1000$ (that is, halo masses greater than $\sim 10^{11}\munit$),
$\Qlim = 0.5$, and $\log_{10}\jsclim = -1.5$.  When considering
changes to the inner halo, the criteria for $\Np$ and $\jsc$ are
replaced by limits on $\Npin$ and $\jscin$, using the same threshold
values.

As in \citetalias{paperI}, we apply two additional selection criteria
suggested by a visual inspection of the time series of properties of
individual haloes.  Firstly, we use a simple measure of ``formation
time'': we restrict our analysis to the time period after the last
time when $M(z) < \Mform$, and choose $\Mform = 0.5 M_0$ (where $M_0$
is the halo mass at $z=0$).  Before this time, haloes tend to have a
much higher rate of accretion and mergers, and experience a general
instability in their properties.  Excluding this period ensures that this does not dominate our results. \pbnew{While this undoubtedly affects the number of major mergers that we expect to see in our sample, we are still interested in how spin orientation changes are distributed amongst mass changes large and small; we're not explicitly excluding post-`formation' major mergers.}  

Secondly, we found that the halo finder and merger tree algorithms
sometimes joined a satellite halo into a larger object as a subhalo,
then separated it off again at the next timestep, perhaps merging
again later.  This will clearly cause large apparent changes to the
halo angular momentum, mimicking a physical spin flip; however it is
due to uncertainty in the halo boundary, rather than a physical change
in the halo angular momentum.  In order to eliminate such ``fake
flips'', we exclude events with large changes in the virial ratio; in
particular, events with $\Delta Q \leq -0.3$ are excluded.  Since this
effect is due to uncertainties in the halo boundary, we do not apply
this exclusion criterion when considering the inner halo spin.

Finally, we note that we analyse our halo population over the redshift
range $z<6.2$; in any case, the effects we describe will be most
observable at low redshift.


\subsection{Evolution of halo properties}
Combining the merger tree data with the halo catalogues at
each timestep means we can obtain the time series of the evolution of
each halo property, for each halo (more precisely: for each halo
\textit{track}, both those that survive until $z=0$ and those that do
not).  As in \citetalias{paperI}, we are most interested in the
changes in halo mass and spin orientation over time.  We therefore
define the same two key quantities used in the previous paper, the
fractional mass change,
\begin{equation}
  \label{e:DMfrac}
  \DM(t) := \frac{M(t) - M(t-\tau)}{M(t)},
\end{equation}
and the change in spin orientation
\begin{equation}
  \label{e:costheta}
  \cos\theta(t) := \frac{      \vv{J}(t)   \dotprod   \vv{J}(t-\tau)}
                        {\left|\vv{J}(t)\right| \left|\vv{J}(t-\tau)\right|}
\end{equation}
where $\tau$ is the timescale over which we measure the halo property
change (the time $t-\tau$ precedes the time $t$).

To allow a fair comparison between spin flips in different sizes of
haloes at different times, we use a constant timescale $\tau$, rather
than simply using the (irregular) time difference between simulation
snapshots.  We linearly interpolate halo properties between snapshots
to get their values at each `previous' time $t-\tau$.  \pbnew{(}The simulation
snapshots are \pbnew{sufficiently well spaced in time that a more complex interpolation scheme is unnecessary)} \pbscrap{spaced closely enough in time that this scheme is
sufficiently accurate.} We choose an event timescale of $\tau =
0.5\Gyr$\pbnew{.}\pbscrap{; the} \pbnew{The snapshot spacing and the impact of our} choice of dynamic timescale \pbnew{are shown} \pbscrap{and its impact
is discussed} in Appendix~\ref{s:timescales}.

We will refer to the halo property changes at a given timestep,
$\DM(t)$ and $\cos\theta(t)$, generically as an \emph{event}.  We
shall use some fiducial values to divide the distribution of events
and aid interpretation: We consider a spin direction change of at
least $\theta_0 = 45\degr$ to be `large', and a fractional mass change
of more than $\DM_0 = 0.3$ to correspond to a major merger.  For the
sake of brevity, we shall often refer to events with $\DM \leq 0.3$ as
minor mergers, even though they could be smooth accretion (i.e. mass
gain from particles that were not from a separate satellite halo), or
even mass loss. Note that the only restriction on $\DM$ is that it
must be below unity.  A value of $\DM = \frac{1}{3}$ corresponds to a
mass gain of $50\%$; our fiducial value of $\DM = 0.3$ results in a
slightly smaller gain of $\frac{3}{7}\approx 43\%$.  If $\DM >
\frac{1}{2}$, then the halo has more than doubled in mass.  We expect
such events to be rare (but not impossible), due to how the merger
trees are constructed: we are always comparing a halo with its most
massive progenitor.  On the other hand, it is also possible for mass
to be lost between timesteps, although again our prejudice is for this
to be rare.  A value of $\DM = -1$ corresponds to a mass loss of
$50\%$.

\pbnew{Finally, note that, as we are focusing on events that can disturb a halo, we are not distinguishing between mergers of two similar-mass haloes and the accretion of multiple small haloes onto a larger one: both cases could be registered as `major mergers' if they occur rapidly enough, such as between two snapshots.  Similarly, we do not consider the direction of halo accretion.  On the other hand, rapid merging of haloes from opposite directions (e.g. along a filament) is likely to result in rather chaotic changes in spin direction -- the infalling haloes would have to have extremely well balanced angular momenta for them to cancel sufficiently for the direction of the vector to remain unchanged, even if their magnitudes nearly cancelled (although the mass gain would still mean that it would be seen as a `disruptive' event).   The frequency of such events would have to be assessed in simulations with higher time resolution, and we note that our results can, in that sense, be seen as a lower limit: more spin flips might be seen in simulations with more timesteps.}

\section{Results}\label{s:results}

\subsection{Flips of the whole halo}

\subsubsection{Distribution of flip and merger events}\label{s:flipdistro}
\begin{figure} 
  \centering\includegraphics[width=\figw]{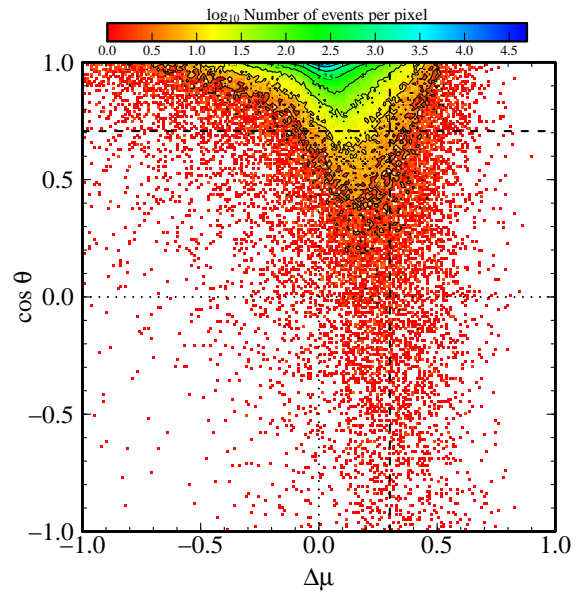}
  \caption{Event distribution in terms of fractional mass change, $\DM$,
    and angular momentum orientation change, $\cos\theta$.  Dotted
    lines mark the origin, and dashed lines indicate our fiducial
    critical values for major mergers ($\DM > 0.3$) and large flips
    ($\theta \geq 45\degr$).  Contours are drawn at the levels given
    by the tickmarks on the colour bar.
 }
  \label{f:distronormalltracks}
\end{figure}

The joint distribution of spin direction changes and fractional mass
changes, for all  \pbnew{$524\,668$} selected events from halo tracks after $z=6.2$, is
shown in Fig.~\ref{f:distronormalltracks} (note that the colour scheme
is logarithmic).  The distribution is very broad, with a strong peak
for events with minimal change ($\cos\theta\approx 1$, $\DM \approx
0$): There are $318\,747$ events ($60.8\%$) just in the range $-0.05 <
\DM \leq 0.05$ \& $\cos\theta \geq 0.95$.  The tail down to larger
spin orientation changes (lower $\cos\theta$) appears biased towards
positive mass change, i.e. mergers.  Note however that many events
have $\DM<0$, i.e. mass \emph{loss} between timesteps. \pbnew{Such events can occur for a number of reasons: just `noise' related to the halo finder and other algorithms, with individual particles being included/excluded from the haloes from one snapshot to the next; and genuine loss from dynamical encounters.  While about $23\%$ of events have $\DM\leq 0$, only $3.7\%$ ($19\,173$) have $\DM\leq -0.1$. Substantial mass loss is even more rare: } \pbscrap{As anticipated
however, low values of $\DM$ are actually very rare: there are very
few events with $\DM\sim -1$, which corresponds to a mass loss of
$50\%$ (} just $340$ events have $\DM\leq -1$\pbnew{, which corresponds to a mass loss of $50\%$.} \pbscrap{).} 



\begin{figure} 
  \centering\includegraphics[width=75mm]{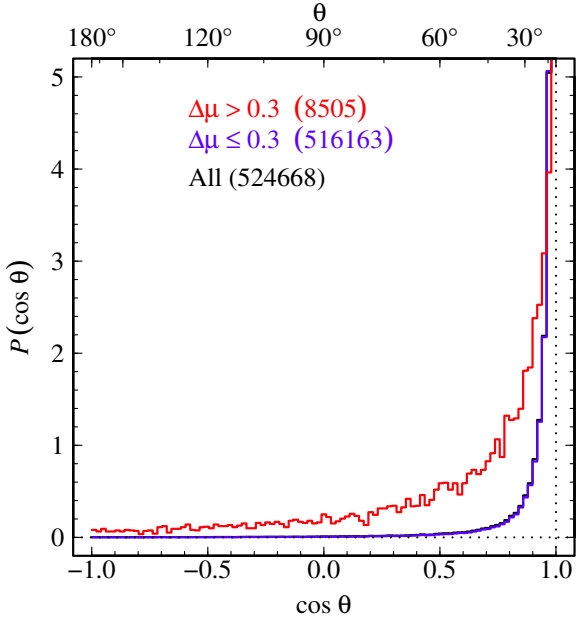}
\caption{Histograms of $\cos\theta$, cross-sections of the event
  distribution shown in Fig.~\ref{f:distronormalltracks}.  The
  histogram of all events is plotted in black, with the subset
  corresponding to major mergers in red and minor mergers in blue. Note that the latter lies almost on top of the black line. The
  number of events selected in each case is written in the
  legend. 
 }
  \label{f:histcos}
\end{figure}

To illustrate the shape of the distribution of selected events more
quantitatively, we now look at cross-sections through it as
histograms.  Fig.~\ref{f:histcos} shows histograms\footnote{Throughout
  this paper, we show histograms normalised like probability density
  functions.} of the spin
orientation change for all halo tracks, plus for the subsections of
the distribution that correspond to major mergers and minor mergers.
The strong spike around ``no change'' is clearly visible, but the
distribution is significantly broader if only major merger events are
considered.

\begin{figure} 
  \centering\includegraphics[width=\figw]{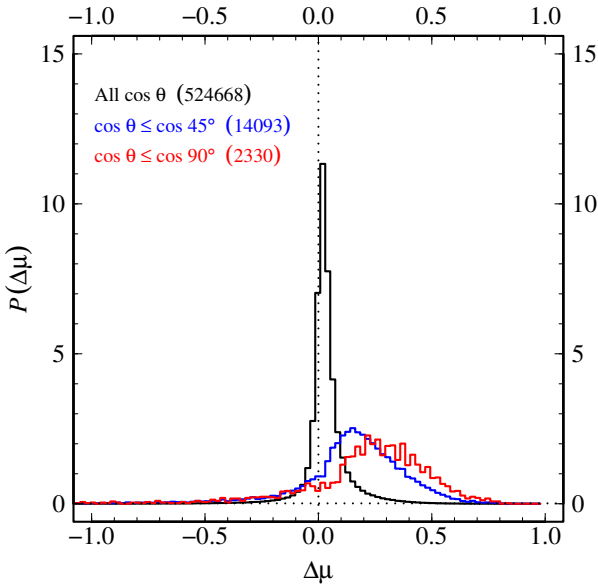}
\caption{Histograms of $\DM$, cross-sections of the event distribution
  shown in Fig.~\ref{f:distronormalltracks}.  The histogram of all
  events is shown in black, while that of the subset of events with
  spin flips of at least $45\degr$ is shown in blue, and for spin
  flips of at least $90\degr$ is shown in red.  The number of events
  selected in each case is written in the legend.
 }
  \label{f:histdmfrac}
\end{figure}

The histogram of halo fractional mass change, for all halo tracks plus
the subsets of those that coincide with spin flips of two different
magnitudes, is shown in Figure \ref{f:histdmfrac}.  Again, the spike
at ``no change'' is clear, but the skewing of the peak of the
distribution towards stronger mergers is also visible, particularly
when the histogram is restricted to events with larger spin
orientation changes.

\begin{figure} 
  \centering\includegraphics[width=\figw]{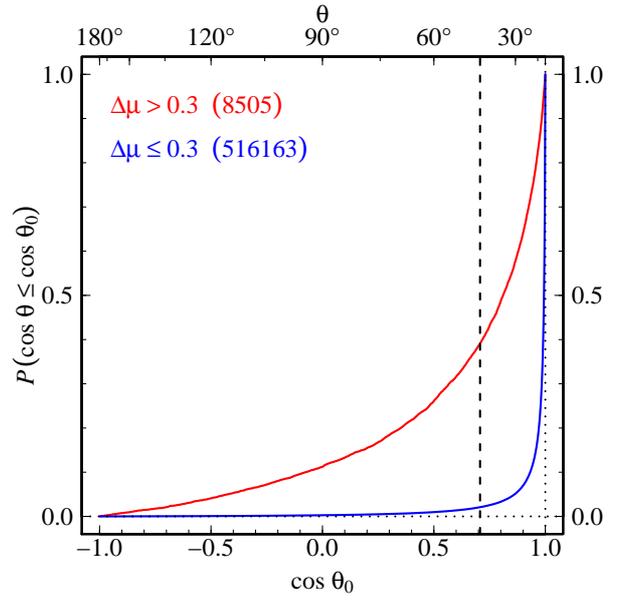}
  \caption{The cumulative distribution of events with spin orientation
    changes of at least $\theta_0$ degrees, from the distribution
    shown in Fig.~\ref{f:distronormalltracks}.  We show results of
    selecting just major merger events ($\DM>0.3$, red) and just minor
    merger events (blue).  Our fiducial value of $\theta \geq 45\degr$
    is marked with a dashed
    line. 
 }
  \label{f:coscdf}
\end{figure}

\begin{figure} 
  \centering\includegraphics[width=\figw]{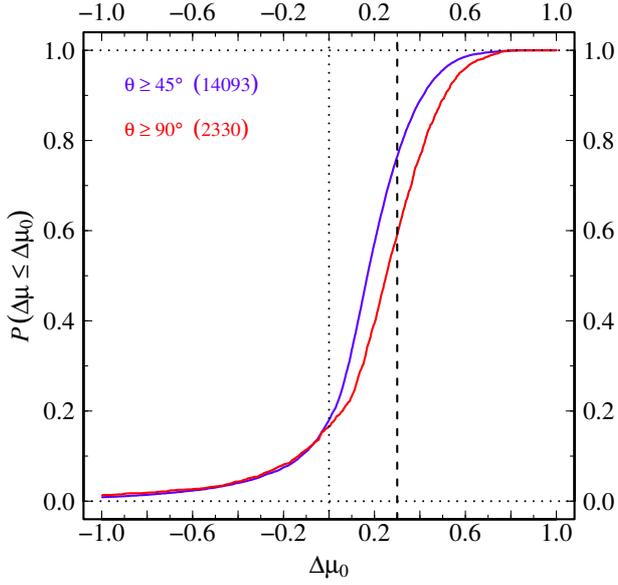}
  \caption{The cumulative distribution of events with fractional mass
    change of $\DM_0$ or less, from the distribution shown in
    Fig.~\ref{f:distronormalltracks}.  We show results of selecting
    just events with spin flips of at least $45\degr$ (blue) and at
    least $90\degr$ (red).  Our fiducial value for major mergers, $\DM
    > 0.3$, is marked with a dashed line.
}
  \label{f:dmfraccdf}
\end{figure}

The cumulative distribution functions (CDFs) give us additional
insight into the relationship between the distributions of spin flips
and halo mass changes.  We show the CDF of $\cos\theta$ in
Fig.~\ref{f:coscdf}.  This shows that minor mergers are very unlikely
to coincide with a large flip: only $2\%$ of events without major
mergers had flips of $45\degr$ or more.  For the major merger events
on the other hand, $39\%$ have spin flips of at least that magnitude.
However, the CDF of $\DM$ (Fig.~\ref{f:dmfraccdf}) shows that $76\%$
of large flips ($45\degr$ or more) coincide with \emph{minor} mergers
($\DM \leq 0.3$).  Although these results at first might seem
contradictory, they are clearly evident in the shape of the
distribution in Fig.~\ref{f:distronormalltracks}, and stem from the
very large number of minor merger events.




\begin{figure} 
  \centering\includegraphics[width=\figw]{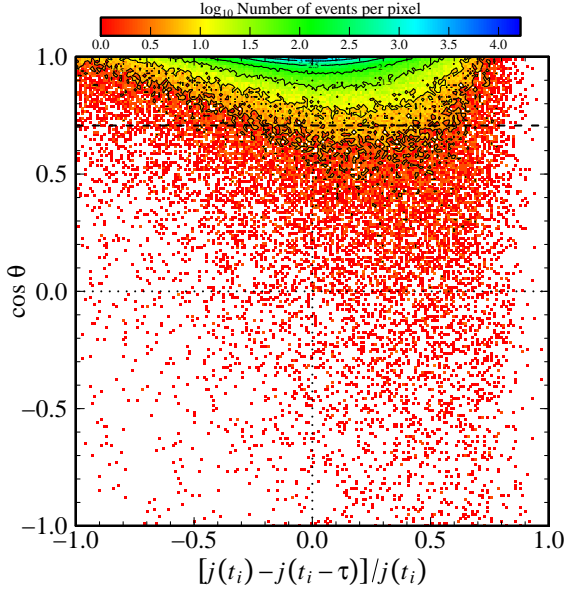}
\caption{Event distribution analogous to
  Fig.~\ref{f:distronormalltracks}, but in terms of the fractional
  specific angular momentum magnitude change instead of the fractional
  mass change ($j(t)=J(t)/M(t)$). 
 }
  \label{f:jdistronormalltracks}
\end{figure}

We show in Fig.~\ref{f:jdistronormalltracks} the
joint distribution of events in terms of the relative change in halo
specific angular momentum \emph{magnitude} versus the
orientation change.  This serves as an important reminder that a spin
flip event does not necessarily mean that the spin magnitude has not
also changed.  Indeed, there is a weak tendency for large spin flips
to correlate with an increase in the halo angular momentum.  There are
still a large number of flips in which the spin magnitude does not
change, and the spin magnitude can change significantly without a
corresponding orientation change.

\subsubsection{Root tracks vs. doomed tracks}
\begin{figure} 
  \centering 
\includegraphics[width=\figw]{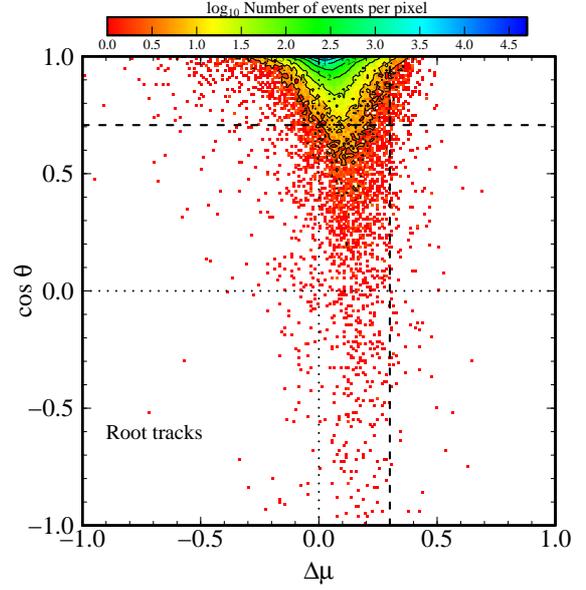}
\includegraphics[width=\figw]{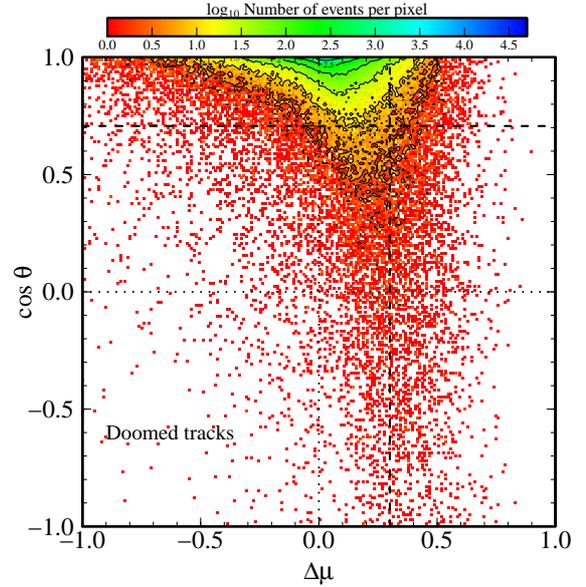}
\caption{The distribution of events, as in
  Fig.~\ref{f:distronormalltracks}, but split into just the $z=0$ root
  tracks only (top) and the doomed tracks only (bottom).
 }
\label{f:distronormsplittracks}
\end{figure}

\begin{figure} 
  \centering 
    \includegraphics[width=\figwtwo]{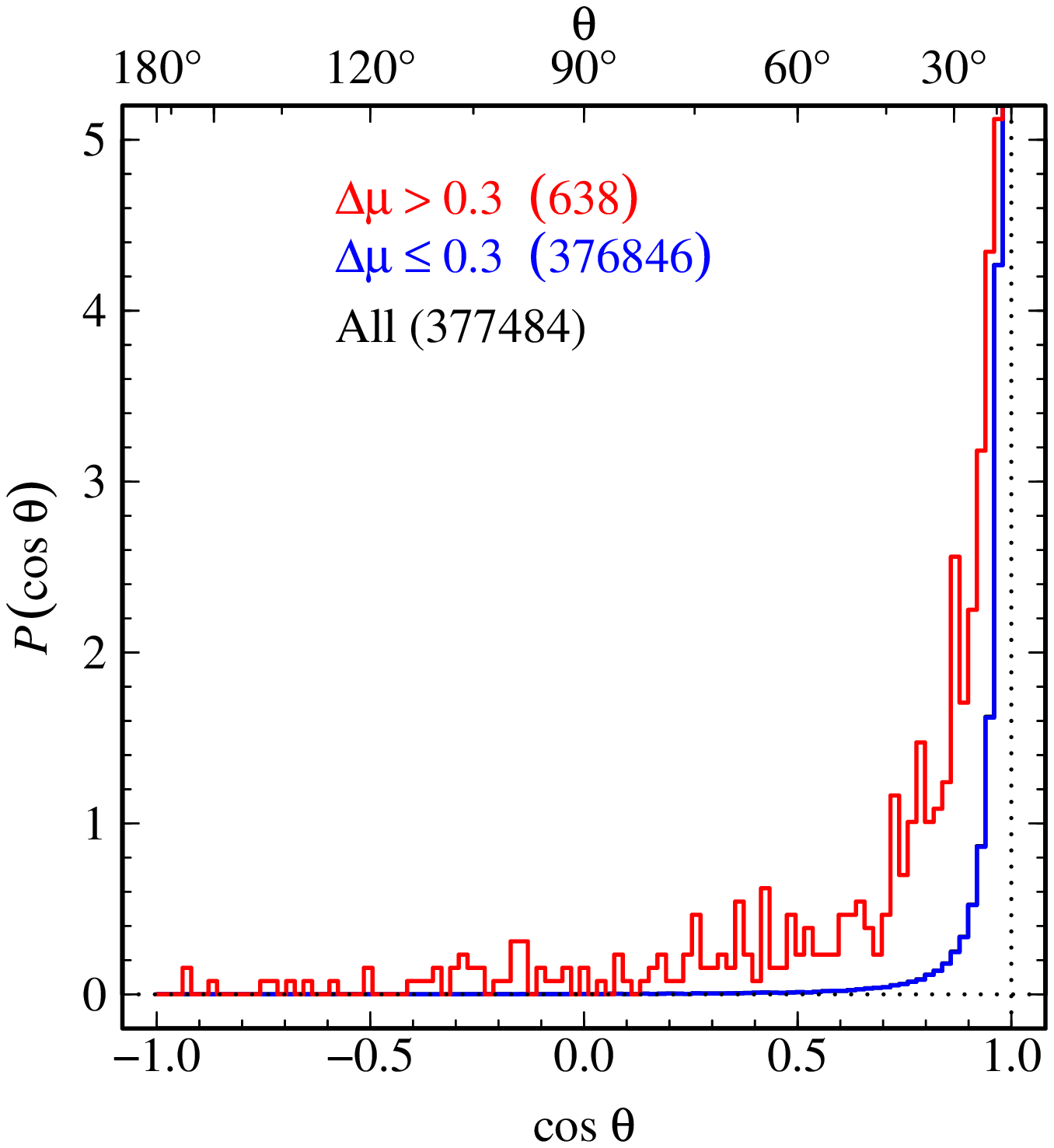}
    \includegraphics[width=\figwtwo]{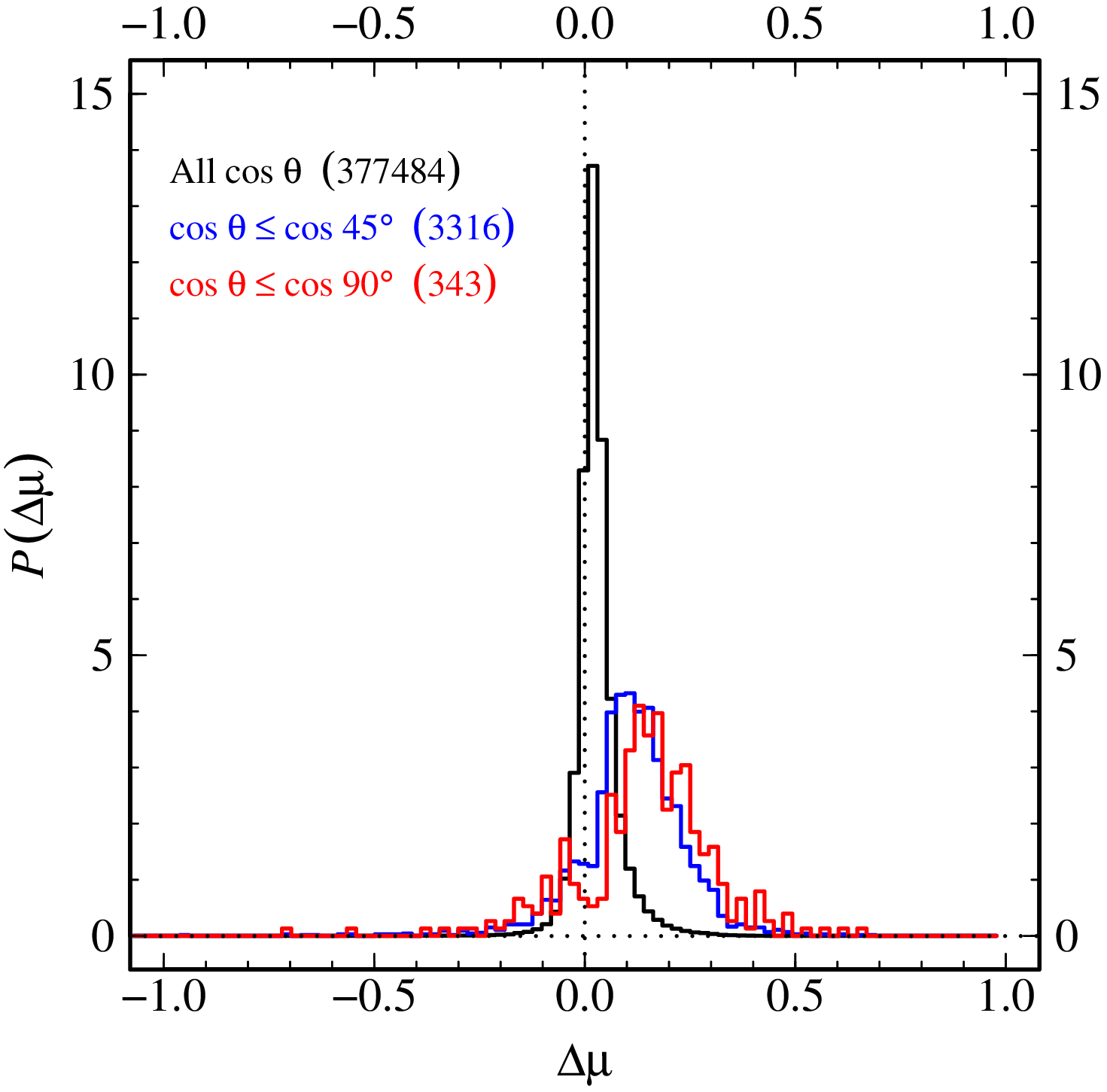}
    \caption{Histograms of the event distribution of root tracks (as
      shown in the top panel of Fig.~\ref{f:distronormsplittracks}),
      in terms of the spin orientation change (top) and the fractional
      mass change (bottom).  These can be compared to
      Figs. \ref{f:histcos} and \ref{f:histdmfrac}.
  }
  \label{f:histssplittracks}
\end{figure}

We split the event distribution shown in
Fig.~\ref{f:distronormalltracks} into that of events in the lives of
the root haloes (those still extant at $z=0$) and events in the lives
of `doomed' haloes (those that merge into a more massive halo before
$z=0$).  The two resulting event distributions are shown in
Fig.~\ref{f:distronormsplittracks}.  Histograms for cross-sections
through the root tracks' distribution \pbnew{of $377\,484$ events} are shown in
Fig.~\ref{f:histssplittracks}.  Although we can see the same basic
trends here as in the previous figures
(Figs.~\ref{f:distronormalltracks}--\ref{f:histdmfrac}), the
distributions are nonetheless noticeably different.  The root
tracks have a much narrower distribution in $\DM$, visible both at
low and high values (mass loss and major mergers).  Much of the broad
tail to more negative $\DM$ seen in Fig.~\ref{f:distronormalltracks}
for all tracks seems to come from the doomed tracks.

\begin{figure}
  \centering
    \includegraphics[width=\figwtwo]{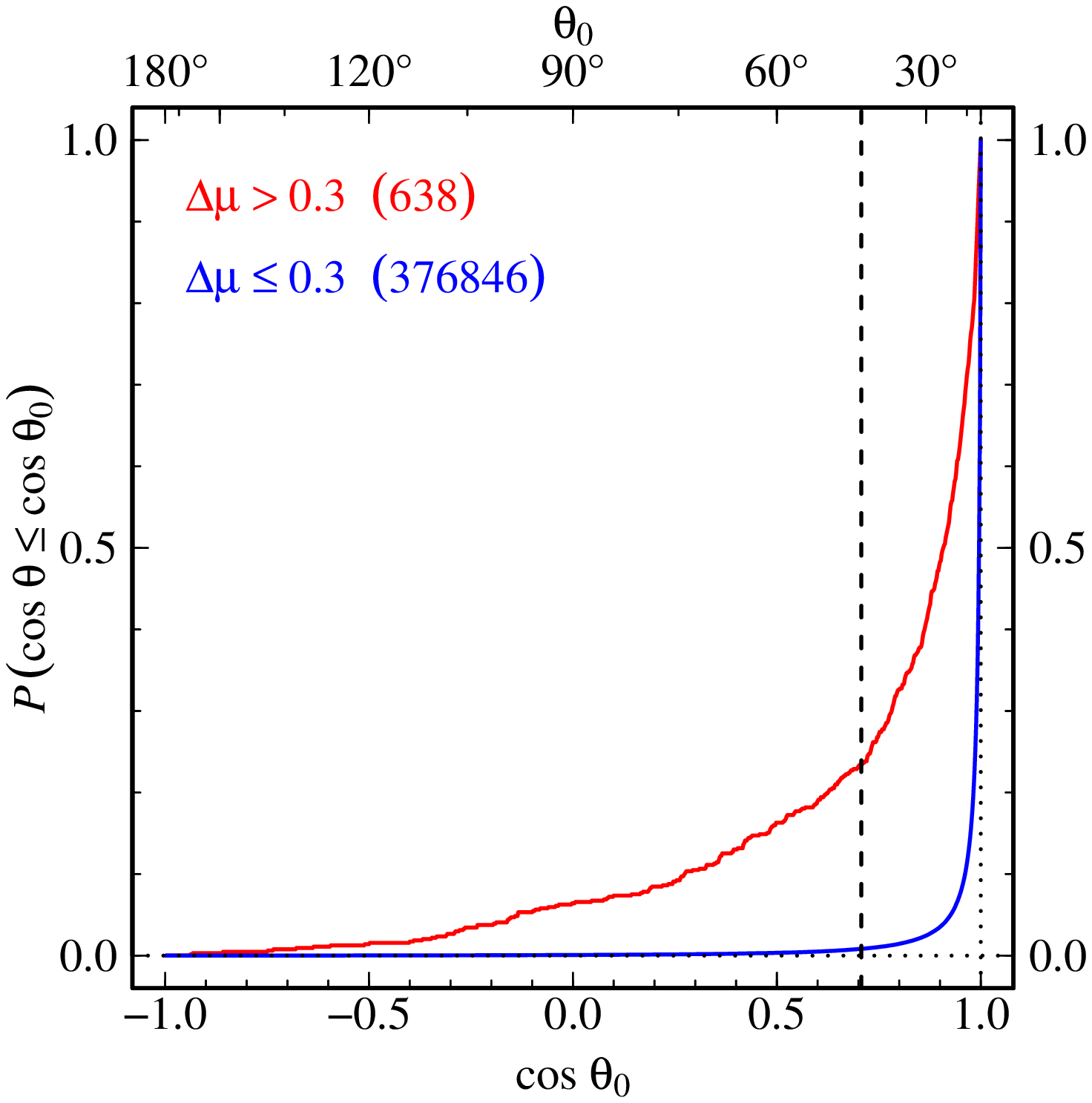}
    \includegraphics[width=\figwtwo]{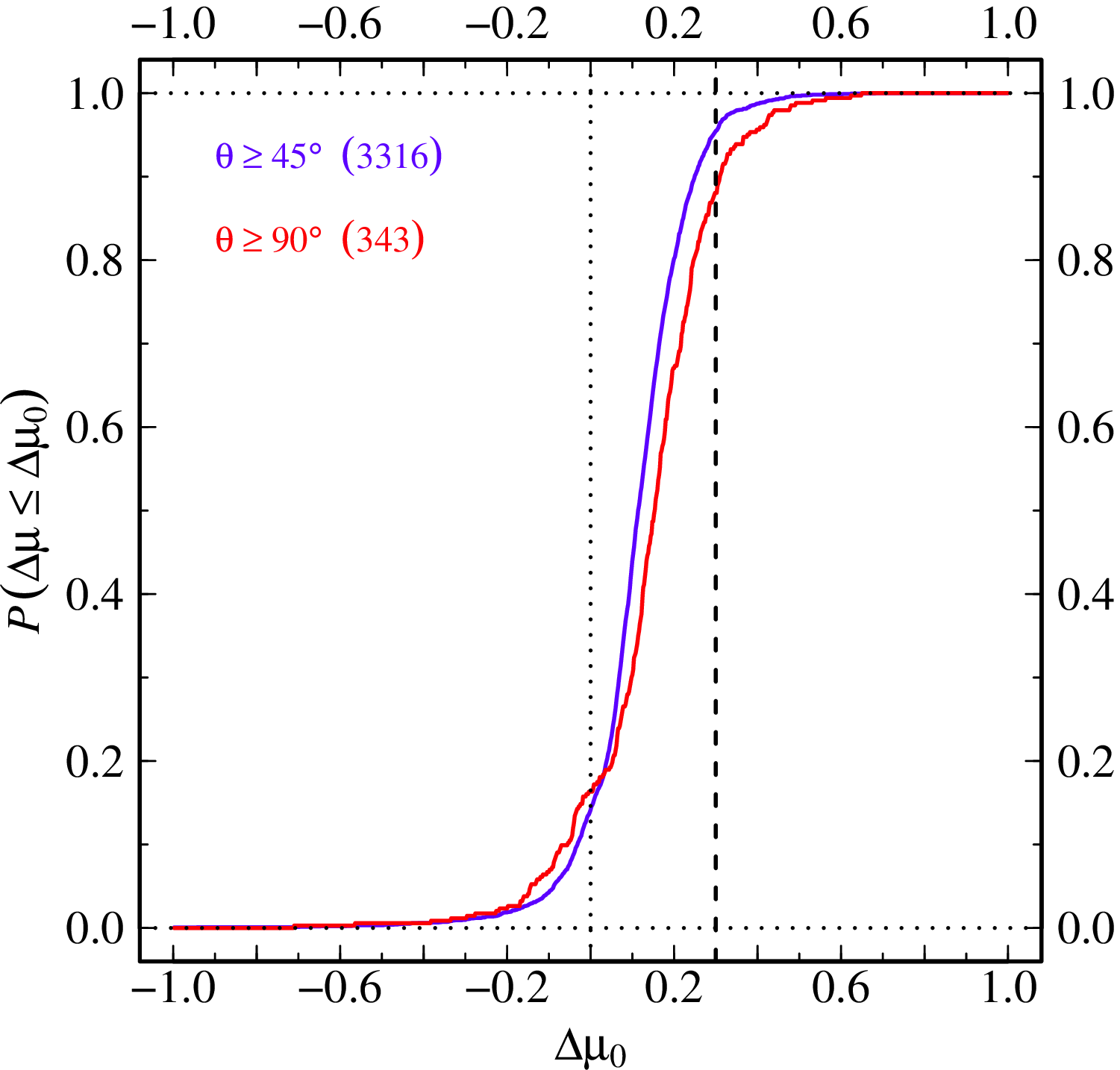}
    \caption{Cumulative distributions of events from haloes on root
      tracks (as shown in Fig.~\ref{f:distronormsplittracks}), as a
      function of flip threshold $\cos\theta_0$ (top) and merger
      threshold $\DM_0$ (bottom).  These can be compared to
      Figs. \ref{f:coscdf} and \ref{f:dmfraccdf}.
  }
  \label{f:cdfsplittracks}
\end{figure}

The cumulative distributions of the events from the root tracks are
shown in Fig.~\ref{f:cdfsplittracks}.  When we consider just these
haloes that survive to $z=0$, we find that less than $1\%$ of minor
merger events have large spin flips, compared to $23.5\%$ of major
mergers.  However, over $95\%$ of spin flips of at least $45\degr$ coincide
with minor mergers ($88\%$ for flips of at least $90\degr$).



\subsection{The inner angular momentum}\label{s:jinner}
\begin{figure*}
  \centering  
  \includegraphics[width=\figwthree]{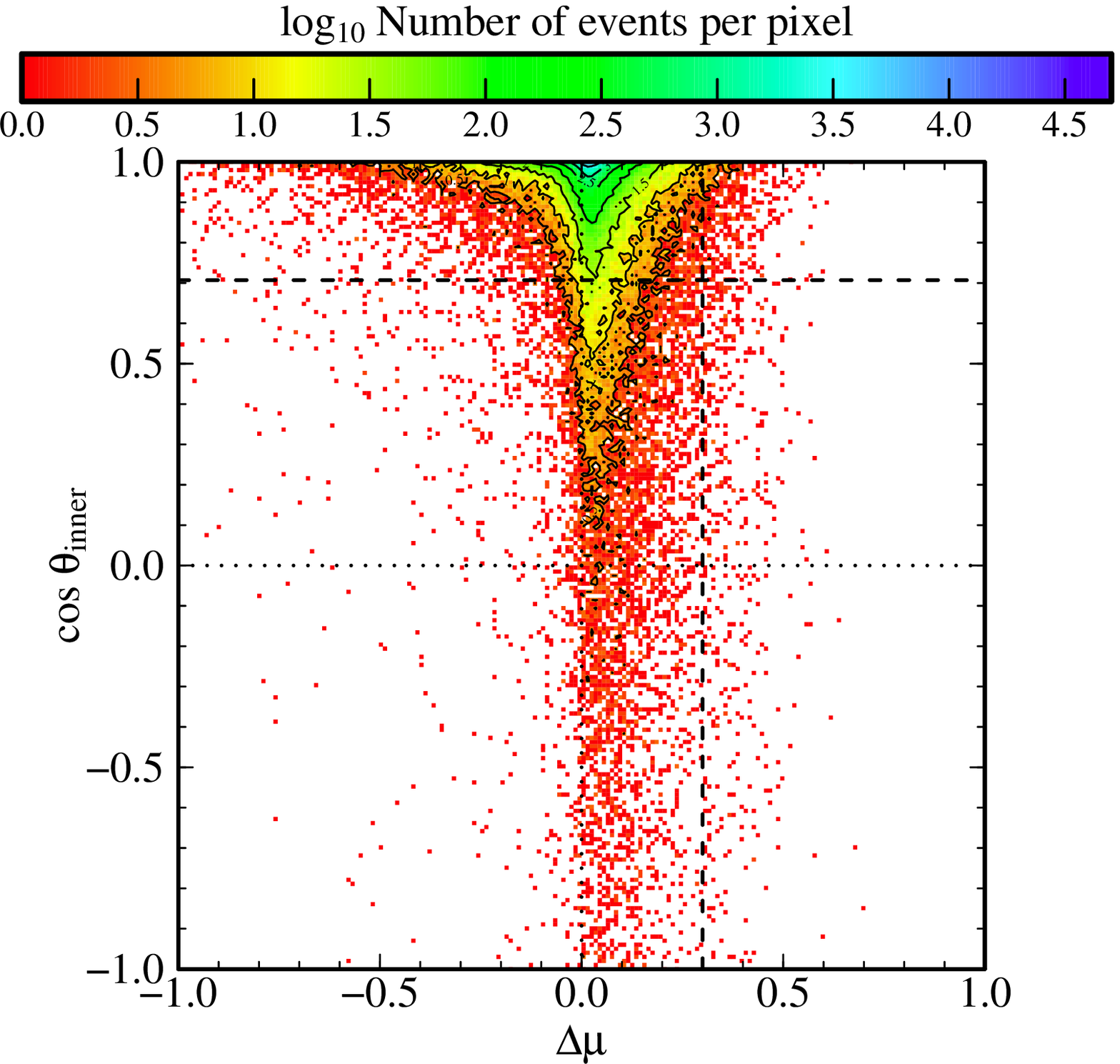}
  \includegraphics[width=\figwthree]{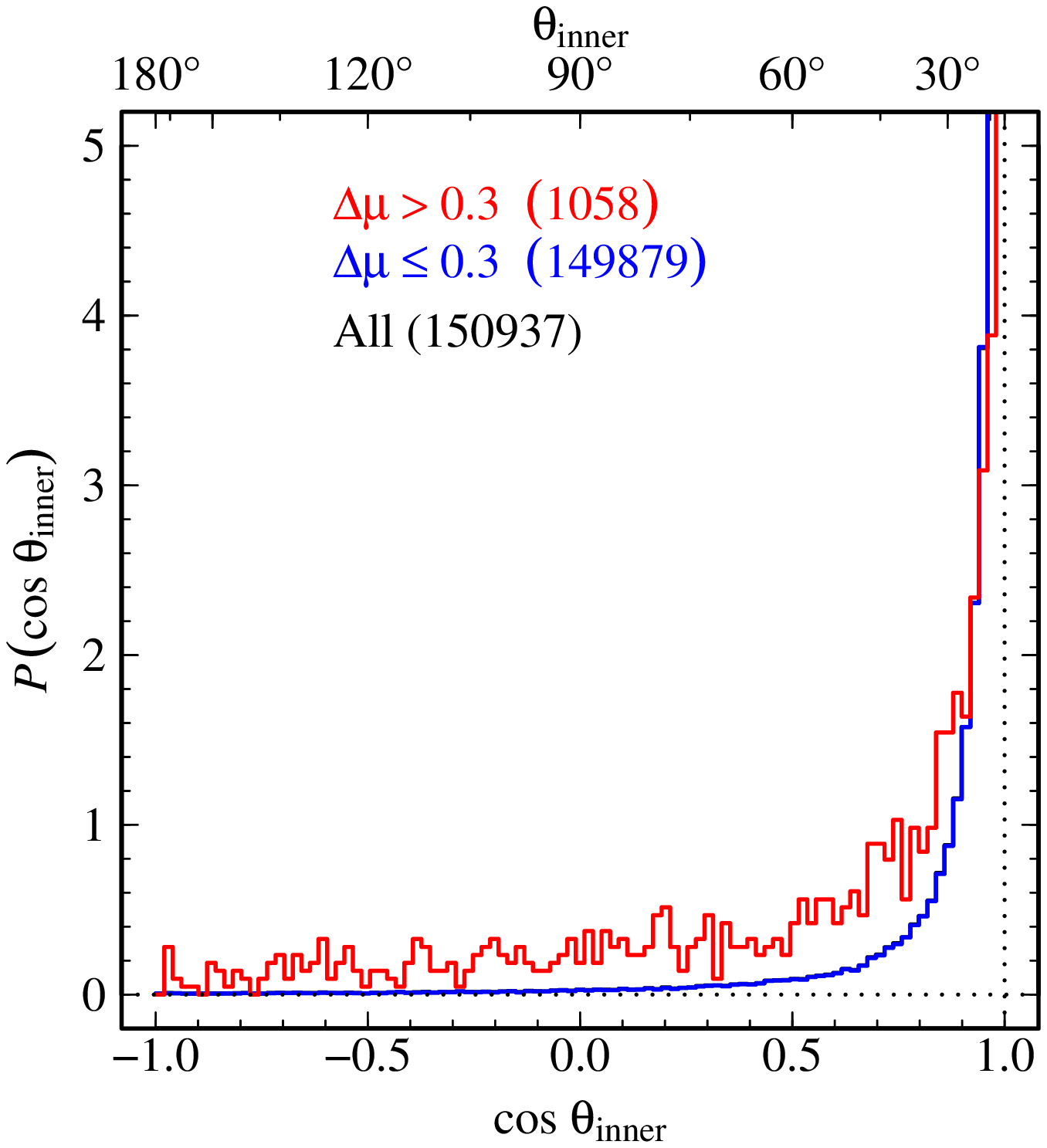}
  \includegraphics[width=\figwthree]{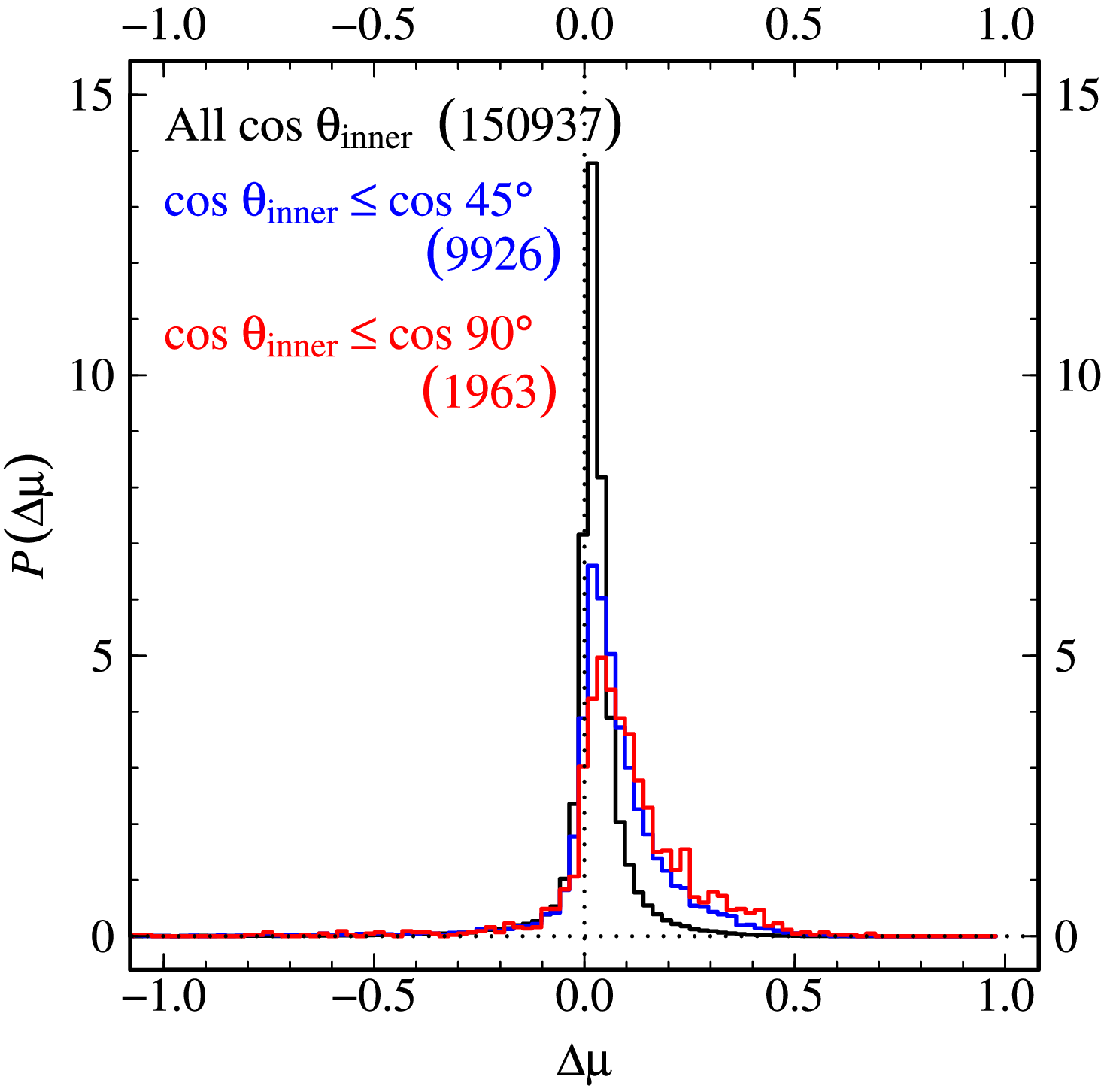}
  \caption{Left: Distribution of events from all halo tracks, in terms
    of the fractional mass change of the total halo, $\DM$, and the spin
    orientation change of the inner halo (i.e. the mass within
    $0.25\Rvir$).  Centre: histogram cross-sections of the event
    distribution, through $\cos\thetain$, for all selected events
    (black) and for the subsets that do (red) and do not (blue) coincide
    with major mergers.  Right: histogram cross-sections through
    $\DM$, for all selected events (black) and for those with spin flips
    of at least $45\degr$ (blue) and $90\degr$ (red).
  }
  \label{f:inneralltracks}
\end{figure*}

To have a noticeable effect on galaxy formation and evolution, it is
reasonable to assume that it is the angular momentum in the
\emph{inner} regions of the halo in particular that needs to change.
We have therefore also looked at the distribution of events from all
halo tracks in terms of the angular momentum of the mass located
within $0.25\Rvir$.  We show the joint distribution of $\cos \thetain$
and the total-halo mass change $\DM$ in the left panel of
Fig.~\ref{f:inneralltracks}, with histogram cross-sections of the
distribution (as before) in the centre and right panels.  There is an
increased likelihood of large spin flips of the inner halo relative to
the total-halo results\pbnew{, although the tighter selection criteria  (i.e. using
$\Npin$ and $\jscin$; see section \ref{s:sel}) means there are fewer events slected overall ($150\,937$)}.  \pbscrap{The modified selection criteria (i.e. using
$\Npin$ and $\jscin$; see section \ref{s:sel})} \pbnew{This also} result\pbnew{s} in far fewer
mass-loss events \pbnew{being selected}.

\begin{figure} 
  \centering\includegraphics[width=\figw]{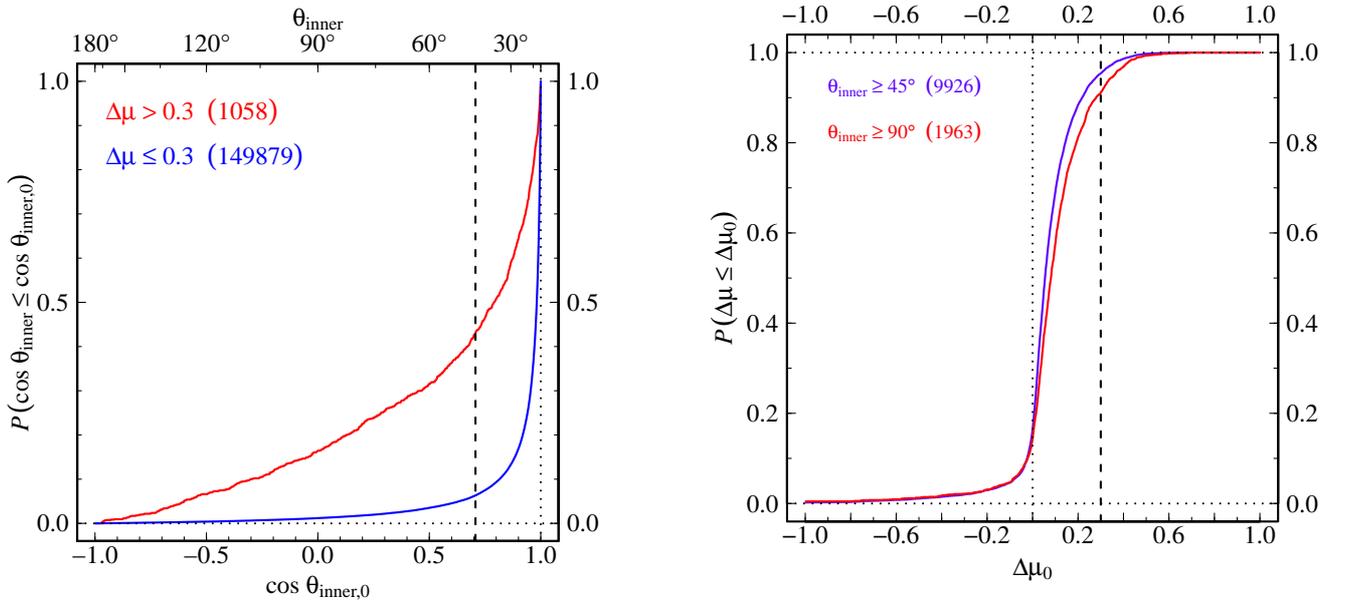}
  \caption{Cumulative distributions of events with inner spin
    misalignments of at least $\theta_{\text{inner,}0}$ degrees.
    Events are selected as in Fig.~\ref{f:inneralltracks}.  We show
    results taking the limiting total-halo fractional mass change
    to be $\DM_0=0.3$ (red: major mergers; blue: minor
    mergers/accretion). 
 }
  \label{f:cosinnercdfalltracks}
\end{figure}

\begin{figure} 
  \centering\includegraphics[width=\figw]{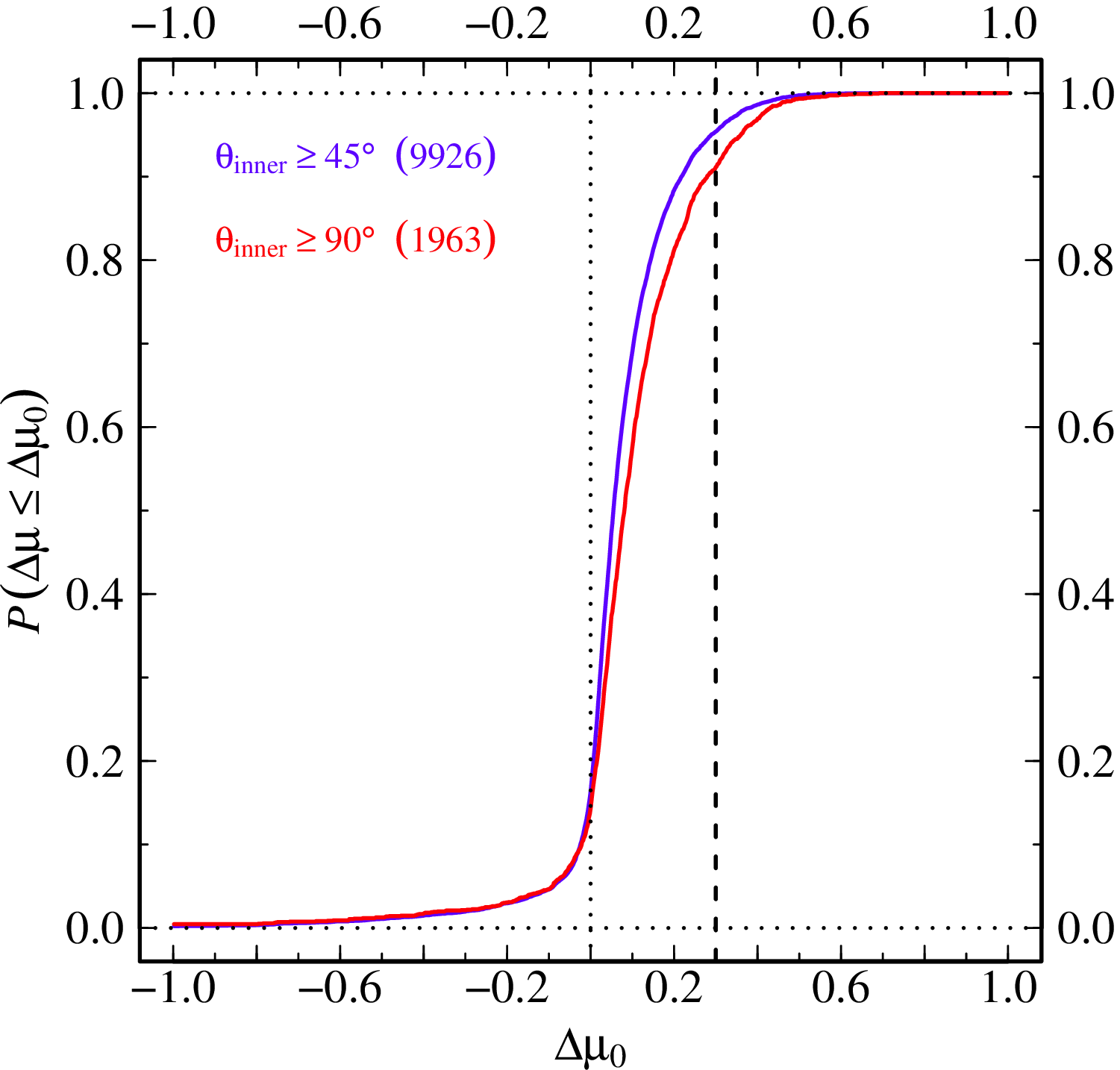}
  \caption{Cumulative distributions of events with fractional mass
    change of $\DM_0$ or less, from the event distribution shown in
    Fig.~\ref{f:inneralltracks}.  We show results for selecting just
    events with inner spin flips of at least $45\degr$ (blue) and at
    least $90\degr$ (red).
 }
  \label{f:dmfraccdfinneralltracks}
\end{figure}

The cumulative distributions for the inner halo spin flip events are
shown in Figs.~\ref{f:cosinnercdfalltracks}
\&~\ref{f:dmfraccdfinneralltracks}.  We find that minor merger events
(the blue line in Fig. \ref{f:cosinnercdfalltracks}) are now much more
likely to coincide with a large inner spin flip: about $6.3\%$,
compared to $0.8\%$ for the total-halo flips shown in
Fig. \ref{f:coscdf}.  The fraction of major merger events that also
have significant inner flips is slightly increased, to $43\%$.  When
selecting just events with large inner flips
(Fig.~\ref{f:dmfraccdfinneralltracks}), we find a strong increase in
the number that coincide with minor mergers: $95.4\%$ for flips of at
least $45\degr$, and $91\%$ for flips of at least $90\degr$.


\begin{figure*}
  \centering
  \includegraphics[width=\figwthree]{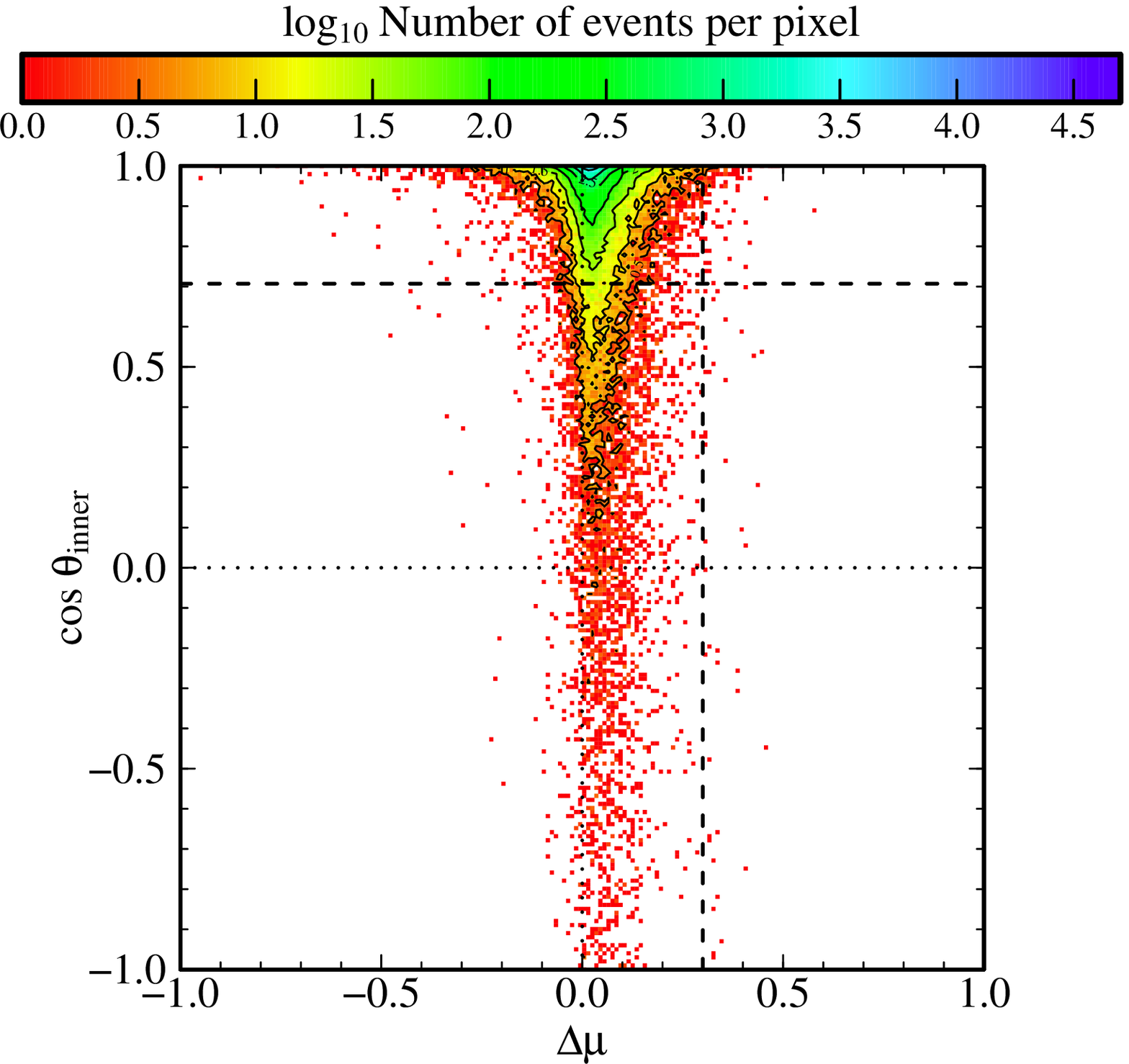}
  \includegraphics[width=\figwthree]{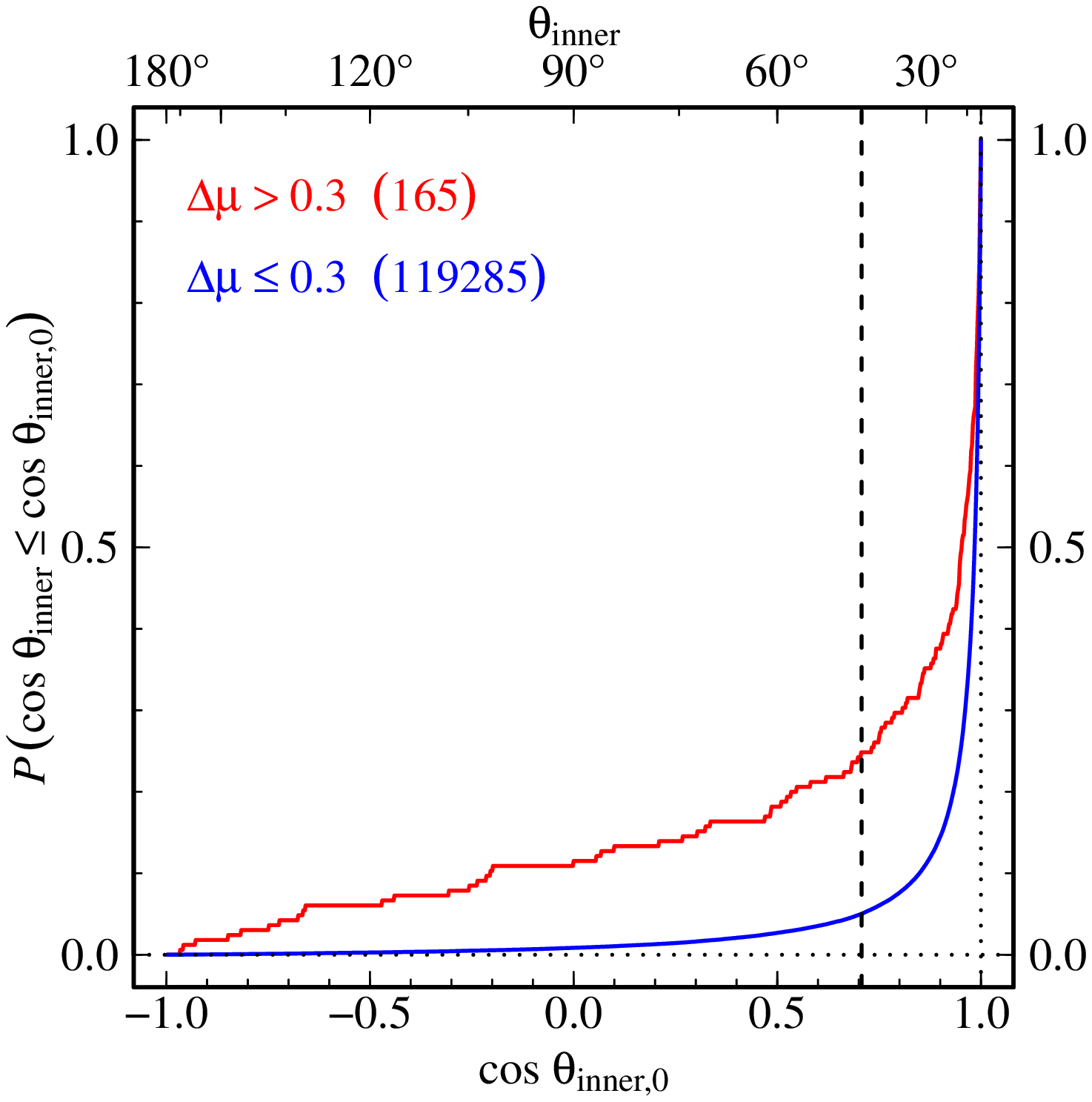}
  \includegraphics[width=\figwthree]{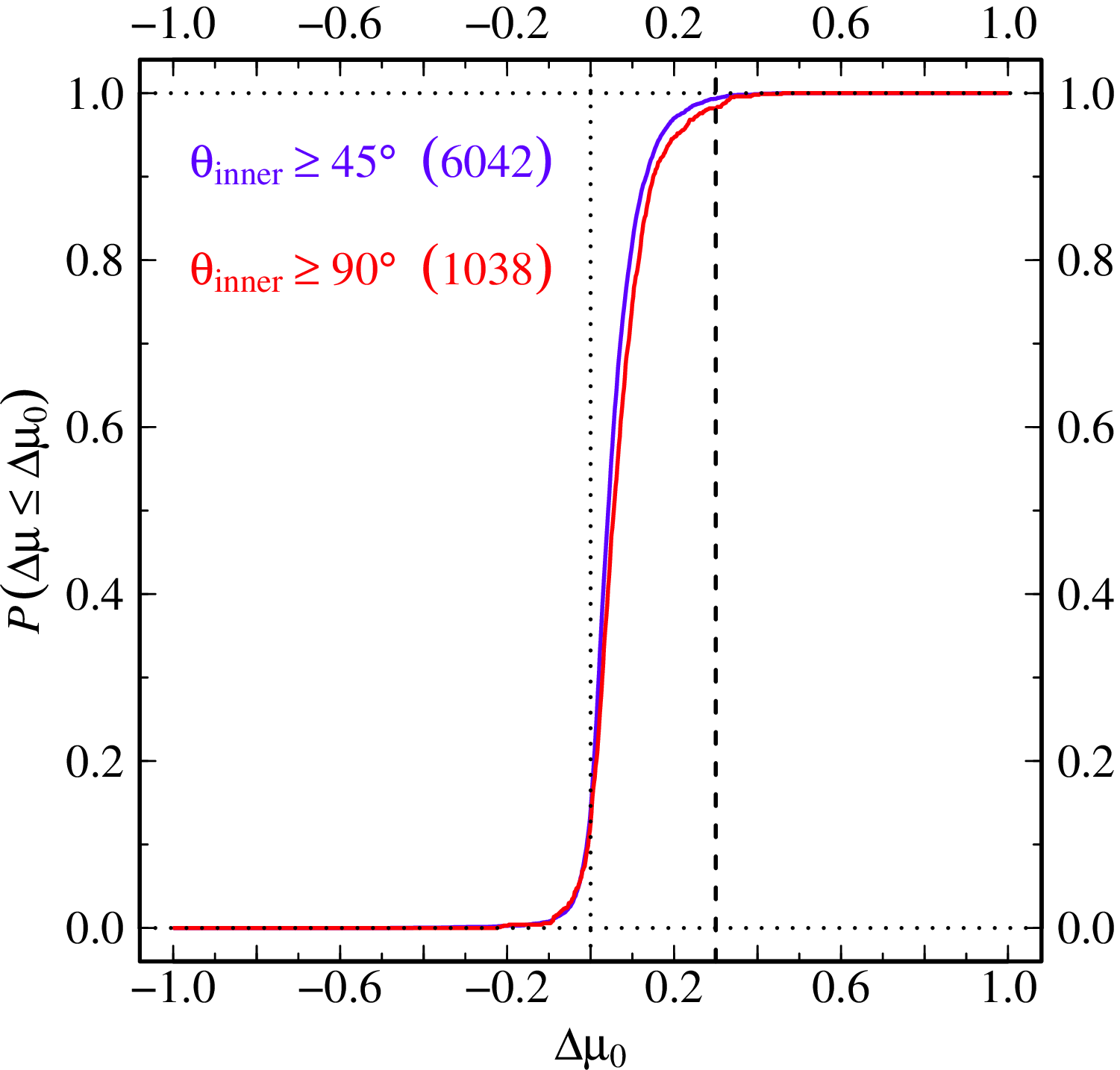}
  \caption{The joint distribution (left) and cumulative distributions
    (centre, right) of events from root haloes in terms of the
    total-halo mass change $\DM$ and inner-halo spin orientation
    change $\cos\thetain$.  These can be compared to
    Figs.~\ref{f:inneralltracks}, \ref{f:cosinnercdfalltracks}
    \&~\ref{f:dmfraccdfinneralltracks}.
  }
  \label{f:distroinnerroottracks}
\end{figure*}

We can also consider the distributions of $\cos\thetain$ and $\DM$ for
just the root tracks (Fig.~\ref{f:distroinnerroottracks}).  Just as
for the total-halo spin, selecting only the root tracks results in a
much narrower distribution of $\DM$.  Although this does not change
the cumulative distribution of minor-mergers much, there are far fewer
large inner flips for the major-merger events (compare the middle
panel with Fig.~\ref{f:cosinnercdfalltracks}).  The fraction of events
with large inner flips that have minor mergers is even higher:
over $99\%$ \pbnew{(of the $119\,450$ selected events)} for flips of $45\degr$ of more.

\subsection{Coincidence of flips and mergers}\label{s:coinc}
It would be na\"ive to assume that, simply because the merger tree
algorithm does not register a major merger occurring at exactly the
same time as a spin flip, the spin flip is not physically associated
with a  major merger.  For example, a major merger could have occurred
at a slightly earlier or slightly later timestep in the simulation.

In fact, a visual inspection of the co-evolution of halo properties
for individual objects suggests that many large spin flips that appear
to coincide with minor mergers actually \emph{do} have a major merger
associated with them, albeit at an earlier or later time.  For
example, the orbit of a satellite halo might take it skimming by the
boundary of a halo for a few timesteps, affecting the dynamics of the
larger halo (causing a flip) before the halo finder deems the
satellite to have actually merged. In the case of changes to the inner
halo, we might anticipate that the inner spin direction would only change some
time \emph{after} a major merger in the halo as a whole, as it might
take some time for the dynamics of the inner region to be affected.
In this section, we attempt a simple assessment of the importance of
such non-coincident flips and mergers\pbnew{, without explicitly considering any  causal connections}.

\begin{figure} 
  \centering\includegraphics[width=\figw]{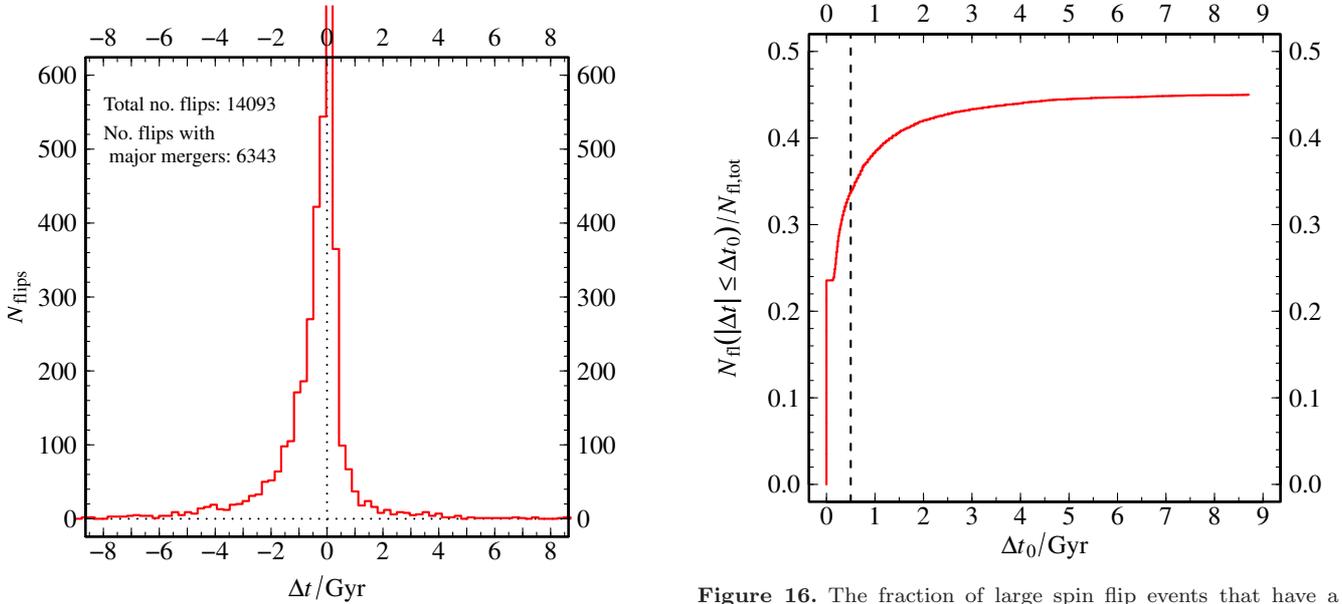}
  \caption{The histogram of the number of spin flip events that have a
    major merger after ($\Delta t>0$) or before ($\Delta t < 0$) the
    flip.  The central peak extends to $N_\text{flips} \sim 3300$.
}
  \label{f:coinchist}
\end{figure}

For each large flip event ($\theta(t_i) \geq 45\degr$) from the
distribution shown in Fig.~\ref{f:distronormalltracks}, we scan the
time series of that halo's fractional mass change both before and
after the time of the flip, $\DM(t_i\pm \Delta t)$.  We record the
time difference between the flip and the nearest major merger
($\DM>0.3$), and plot a histogram of these times, $\Delta t$, in
Fig.~\ref{f:coinchist}.  We see that only $45\%$ of these flips
(i.e. $6343/14\,093$) have major mergers that can be identified at
\emph{any} time, before or after the flip.  For the remaining flip
events, no major merger event could be found within the lifetime of
the halo.  We find that most flips with non-coincident major mergers
have the merger preceding the flip, i.e. the angular momentum vector
swings round \emph{after} the mass has been incorporated: of the flips
with major mergers identified at some point in their lifetimes, there
are $2232$ with $\Delta t<0$, versus $790$ with $\Delta t>0$.


\begin{figure} 
  \centering\includegraphics[width=\figw]{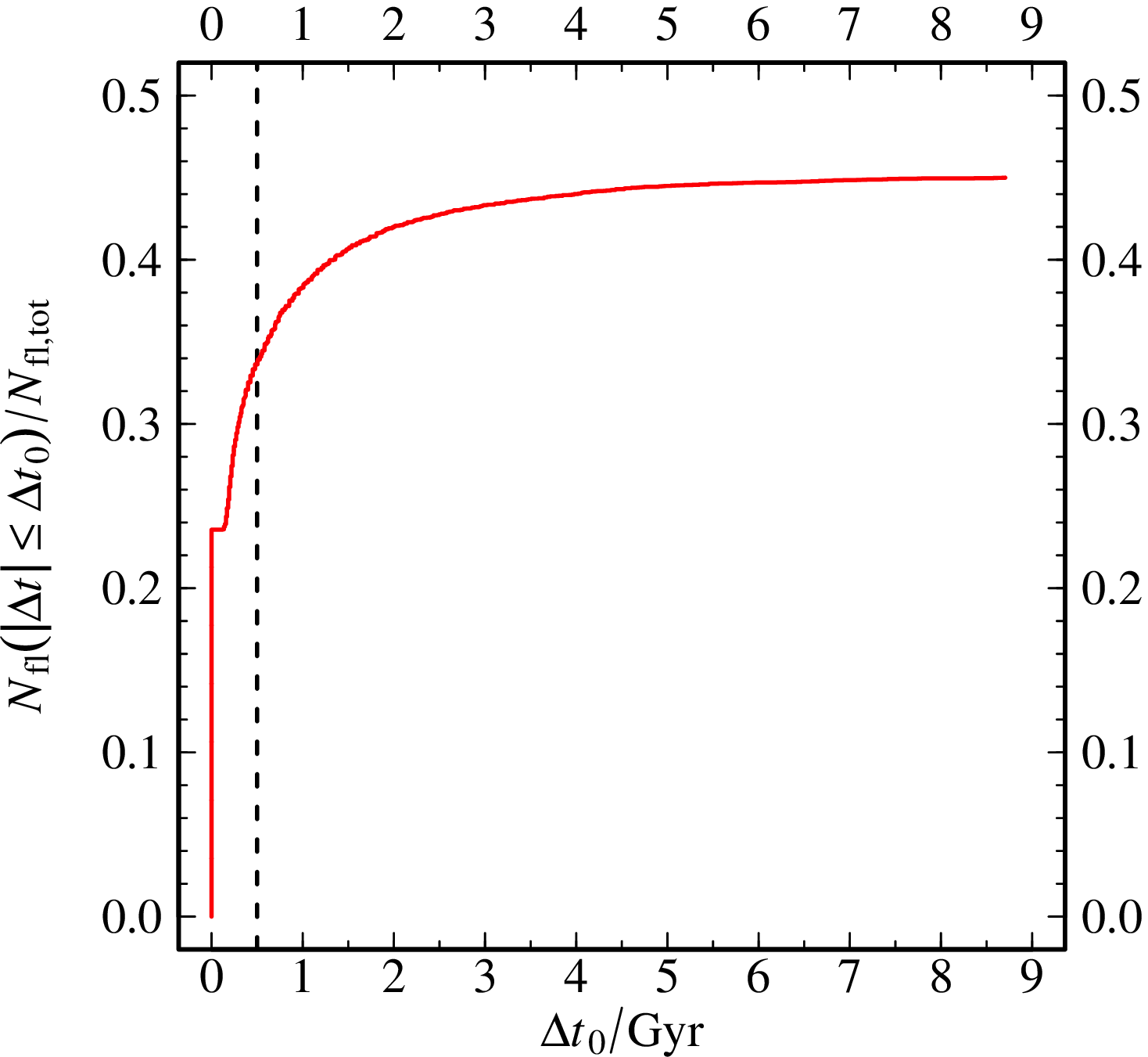}
  \caption{The fraction of large spin flip events that have a major
    merger within a time $|\Delta t| \leq \Delta t_0$.  Our dynamical
    timescale value of $\tau = 0.5\Gyr$ is marked with a dashed line.
}
  \label{f:coinccuml}
\end{figure}

The cumulative distribution is shown in Fig~\ref{f:coinccuml}.  This
shows clearly that $23.6\%$ of the large flip events ($3320$) coincide
``exactly''\footnote{i.e. over the same timesteps in the simulation.}
with major mergers (also seen in Figs.~\ref{f:coscdf}
\&~\ref{f:dmfraccdf}).  The number of coincident major mergers
initially rises very steeply with $\Delta t$ however.  If we allow for
major mergers within $\pm 0.5\Gyr$, then the fraction of large flips
coinciding with major mergers rises to $33.6\%$, and then to $38.3\%$
if we extend the window to $\pm 1\Gyr$.  Beyond this, the CDF grows
only slowly, until reaching the maximum at $6343$ flips ($45\%$).


\begin{figure} 
  \centering\includegraphics[width=\figw]{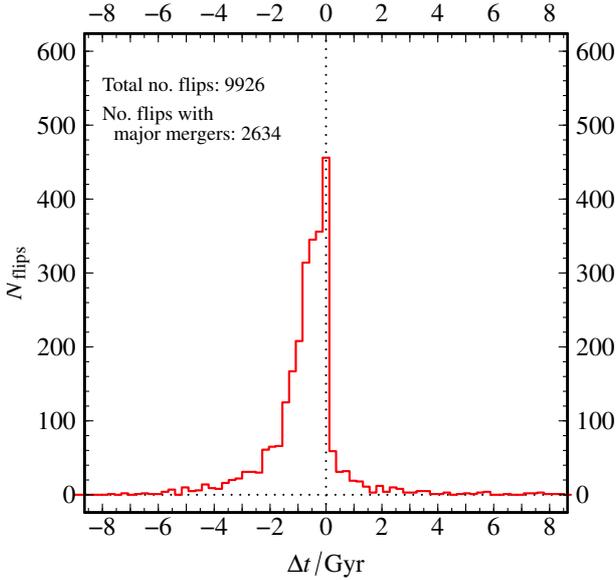}
  \caption{Number of large flips of the inner halo angular momentum
    that have a (total-halo) major merger a certain amount of time
    after ($\Delta t>0$) or before ($\Delta t<0$) the flip.
}
  \label{f:coinchistinner}
\end{figure}

\begin{figure} 
  \centering\includegraphics[width=\figw]{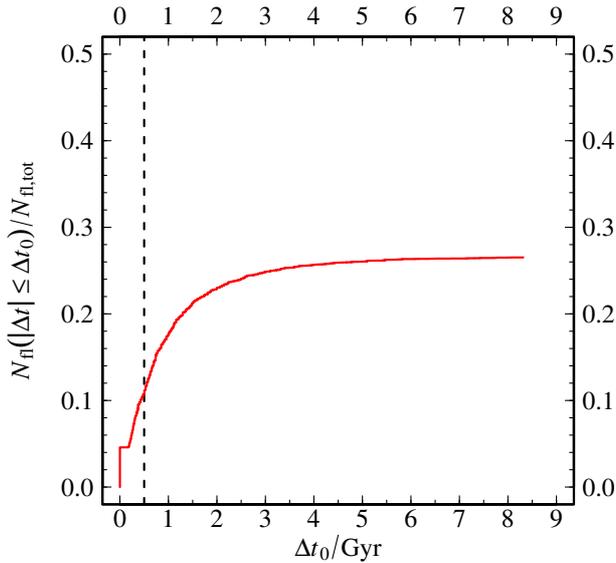}
  \caption{The fraction of inner-halo flip events that have a
    total-halo major merger within a time $|\Delta t| \leq \Delta
    t_0$.
}
  \label{f:coinccumlinner}
\end{figure}

We analyse the inner flips in the same way
(Figs.~\ref{f:coinchistinner} \&~\ref{f:coinccumlinner}),
based on the event distribution shown in Fig.~\ref{f:inneralltracks}.  The
histogram shows that in this case there are far fewer major mergers
following large inner spin flips: $73\%$ ($1928$) of the inner flips associated
with a major merger at any time are preceded by the merger, compared
to $250$ that are followed by the merger. This is what we would
expect: a merger would be initially seen for the halo as a
whole, with the inner halo dynamics reacting a short time later.  From
the cumulative plot (Fig.~\ref{f:coinccumlinner}), we can see that
just $26.5\%$ of large inner flips have total-halo major mergers at
any time before or after.  As can also be seen in
Figs.~\ref{f:cosinnercdfalltracks} \&~\ref{f:dmfraccdfinneralltracks},
only $4.6\%$ of flips ($455$) coincide exactly with a major merger.
If we allow for a time lag between a major merger and the inner halo
spin direction changing, then we find $10.7\%$ of flips ($1065$
events) have a merger within $\pm 0.5\Gyr$, rising to $17.6\%$ ($1745$
events) within $\pm 1\Gyr$.


\subsection{Spin flips over halo lifetimes}\label{s:hflipfracDt}
As in \citetalias{paperI}, we now move on to investigate the
likelihood of large spin flips occurring over the lifetimes of haloes,
rather than the simple distribution of events discussed in the
previous sections.  We wish to find the probability of a halo
undergoing a spin flip of a given magnitude ($\theta_0$) and duration
($\tau$, the event timescale), at some point during its lifetime
(excluding events at timesteps when the angular momentum measurement
is not reliable).  We can also divide this into spin flips during
which the halo's mass does or does not grow by a certain amount,
$\DM_0$ (i.e. considering just flips that do or do not coincide with
major mergers).

\begin{figure*} 
  \centering\includegraphics[width=\textwidth]{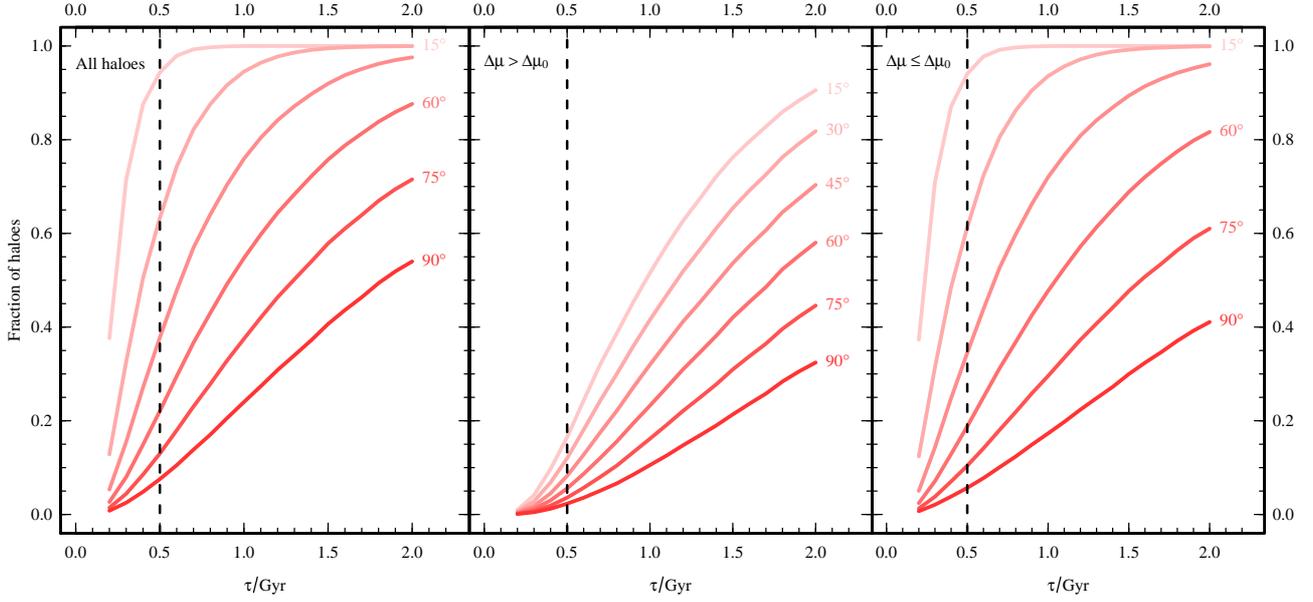}
  \caption{Left: fraction of all halo evolutionary tracks that have at
    least one spin flip of duration $\tau$ and size
    $\geq\theta_0$. Six values of $\theta_0$ have been chosen, every
    $15\degr$ from $15\degr$ to $90\degr$.  Middle: fraction of halo
    tracks with at least one spin flip that coincides with a major
    merger.  Right: same, but for spin flips that do not coincide with
    a major merger.  The characteristic timescale used in the rest of the paper,
    $\tau = 0.5\Gyr$, is marked with a dashed line.
}
  \label{f:htotflipfracDt}
\end{figure*}

We show the results for this in Fig.~\ref{f:htotflipfracDt}.  As one
might expect, as larger timescales $\tau$ are considered, the
likelihood of a halo exhibiting a flip of any given size increases.
For our fiducial values of $\theta_0=45\degr$ and $\tau=0.5\Gyr$, we
find that $37.8\%$ of \pbnew{the $40\,559$ selected} haloes exhibit such a flip at some point in
their lives.  If we consider just those flips that coincide (exactly)
with major mergers, this figure is much lower, at $8.3\%$ (see middle
panel).  Considering just flips that do \emph{not} coincide with major
mergers (right panel), we find that $34.5\%$ of haloes experience such
flips\footnote{Note that the values for ``All haloes'' need not be the
  sum of those whose flips do and do not coincide with major
  mergers. A halo can have flips of both kinds during its lifetime, so
  the categories are not mutually exclusive.}.


\begin{figure*} 
  \centering\includegraphics[width=\textwidth]{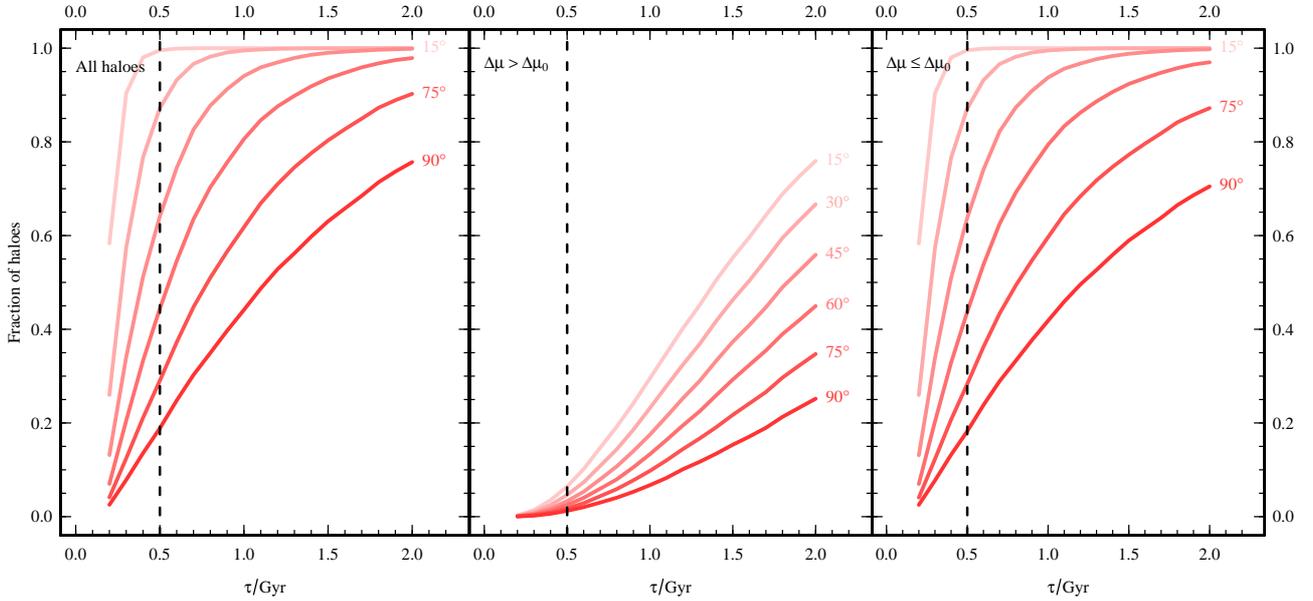}
  \caption{Fraction of all halo tracks that have at least one inner
    spin flip of at least $\theta_0$ and duration $\tau$ (left), and
    the fractions that have such an inner flip that coincides with a
    major merger (middle) or does not (right).  The same six values of
    $\theta_0$ are used as in the previous figure.  The charactersitic timescale used
    in the rest of the paper, $\tau = 0.5\Gyr$, is marked with a
    dashed line. 
}
  \label{f:hinnflipfracDt}
\end{figure*}

We perform the same analysis for flips of the inner halo angular
momentum in Fig.~\ref{f:hinnflipfracDt}.  In this case, haloes are
more likely to experience inner spin flips of any magnitude and any
duration: $64.1\%$ of \pbnew{the $12\,171$ selected} haloes have inner flips of at least $45\degr$
over $0.5 \Gyr$ during the course of their lifetime.  As we have seen,
fewer inner halo flips coincide with major mergers
(e.g. Fig.~\ref{f:coinccumlinner}).  It is therefore not surprising
that, when considering just inner flips that coincide with major
mergers, we get a lower fraction of haloes ($3.3\%$) compared to that
for the total halo angular momentum.


\begin{figure*} 
  \centering\includegraphics[width=\textwidth]{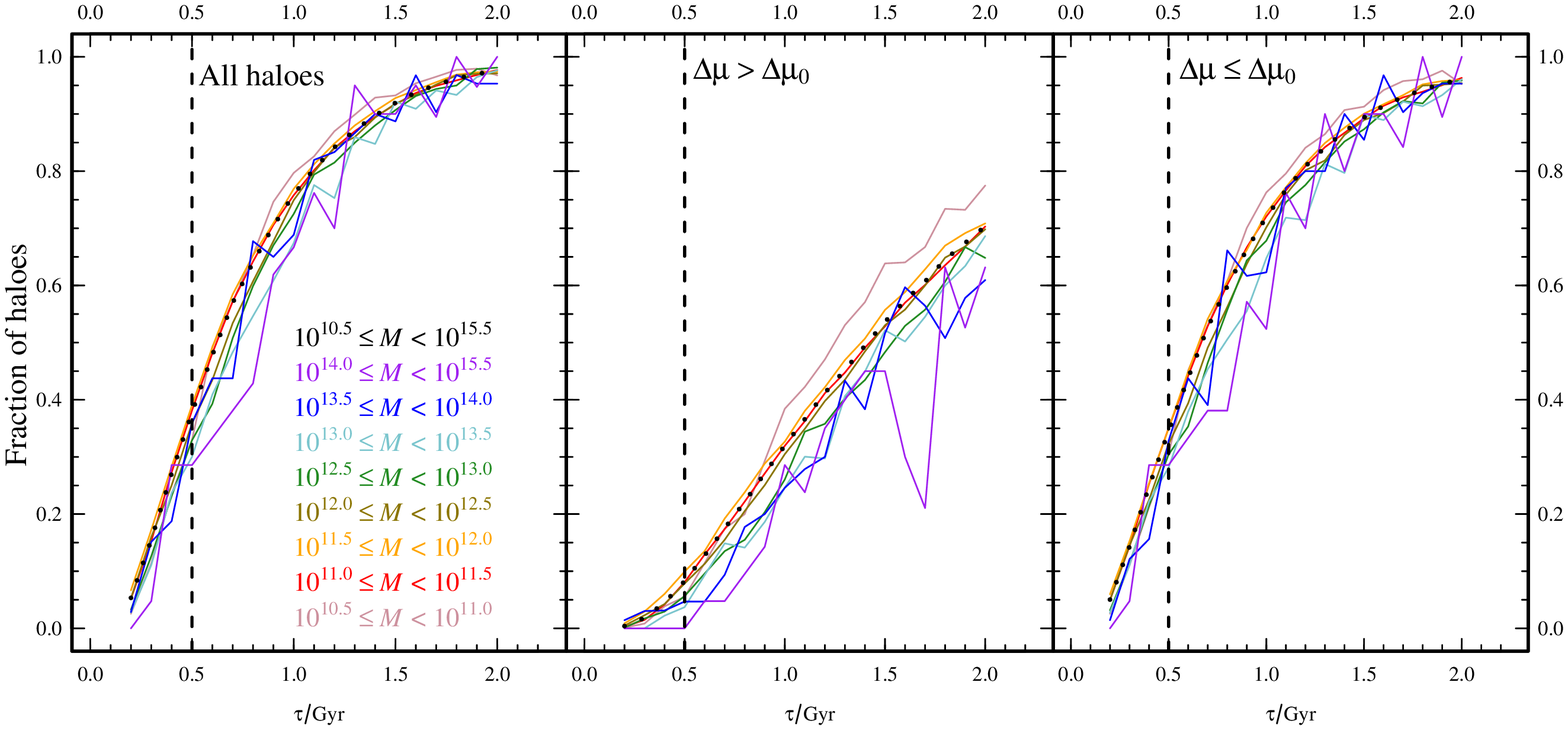}
  \caption{Left: fraction of all halo evolutionary tracks that have at
    least one spin flip of duration $\tau$ and amplitude $\geq
    45\degr$. The middle and right panels also require that the flip
    does or does not coincide with a major merger (as in
    Fig.~\ref{f:htotflipfracDt}).  The haloes are divided into bins
    according to their final-time mass, colour-coded as shown (the
    masses in the legend are in $\munit$). An additional dotted heavy
    black line shows the results over the whole mass range (this is
    the same as the $45\degr$ line in
    Fig.~\ref{f:htotflipfracDt}). Note that the content of the lowest
    mass bin is affected by our particle-number limit for halo
    selection: $1000\mp \approx 10^{10.98}\munit$, and the bin upper
    limit of $10^{11}\munit$ corresponds to about $1050$ particles.
    The timescale used in the rest of the paper, $\tau = 0.5\Gyr$, is
    marked with a dashed line. 
}
  \label{f:htotflipfracDtMbins}
\end{figure*}

Since we have haloes over a wide range of final-time masses (both for
those that survive to $z=0$ and the `doomed' haloes), it is interesting
to ask whether there is any mass-dependence in the probability of
flips over halo a lifetime.  In Fig.~\ref{f:htotflipfracDtMbins}, we
show this, in a plot analogous to Fig.~\ref{f:htotflipfracDt}.  In
this case, instead of plotting results for different spin flip sizes,
we set $\theta_0=45\degr$ and give the results for different
bins of final halo mass.

\begin{figure*} 
  \centering\includegraphics[width=\textwidth]{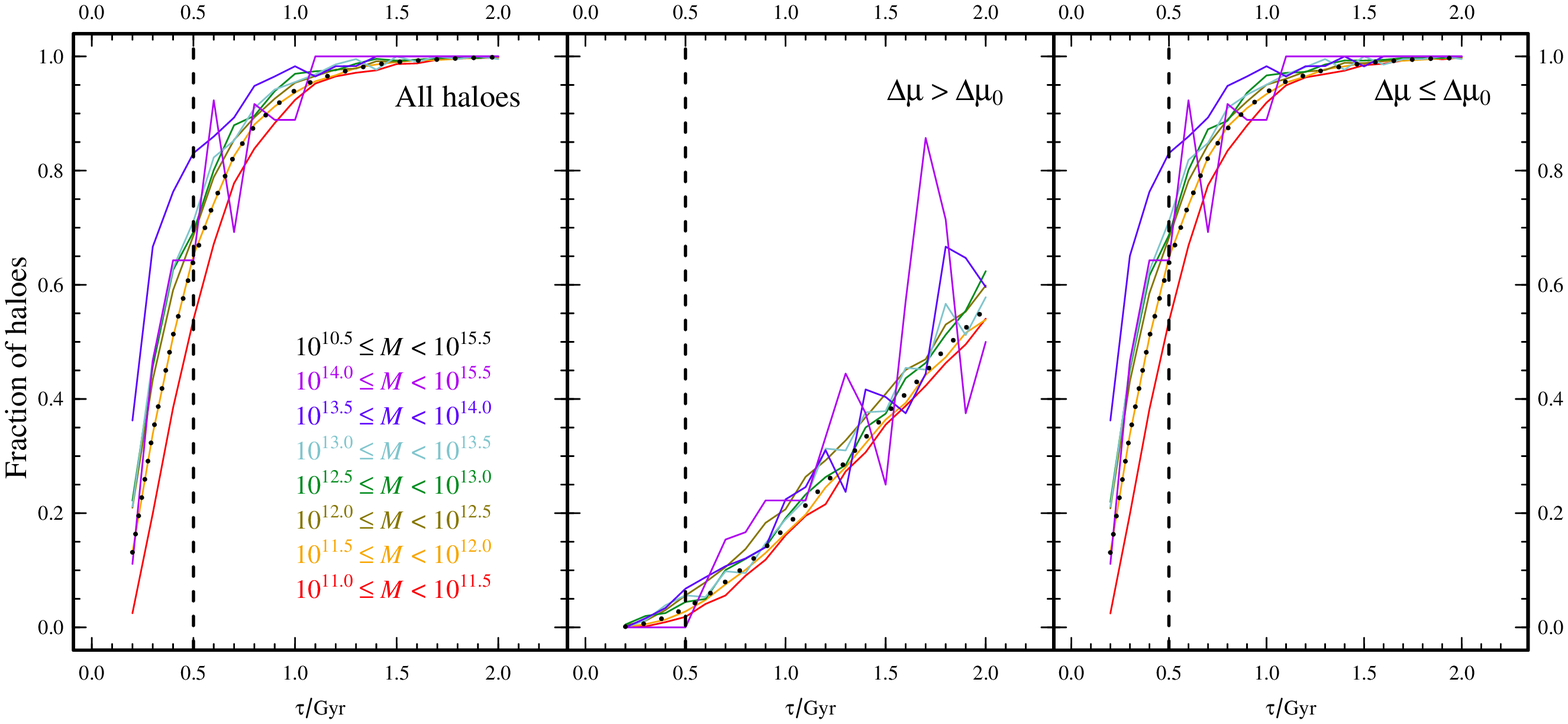}
  \caption{As Fig.~\ref{f:htotflipfracDtMbins}, but considering flips
    in the inner halo spin (left), which do (middle) or do not (right)
    coincide with major mergers.  The bins in total-halo final time
    mass are colour-coded as shown (the masses in the legend are in
    $\munit$), with an additional heavy black dotted line showing the
    results over the whole mass range (as in
    Fig.~\ref{f:hinnflipfracDt}).
 }
  \label{f:hinnflipfracDtMbins}
\end{figure*}

There is a hint of mass-dependence in the probability for flips:
lower-mass haloes appear slightly more likely to experience a flip of
a given timescale during their lifetime.  This trend is reversed when
considering flips of the inner halo angular momentum, shown in
Fig.~\ref{f:hinnflipfracDtMbins}.  In this case, haloes with larger
masses at their final timestep are more likely than lower-mass haloes
to have experienced an inner spin flip at some point in their
lifetimes.  However, in both the total-halo and inner-halo cases, the
differences between different mass bins are very slight, and the
results from the high mass bins in particular are noisy because they
contain relatively few haloes.  What trend there is appears
stronger when looking at inner flips of short duration, but is broadly
consistent over all values of $\tau$ for the total-halo spin flips.

\subsection{The contribution of other progenitors}\label{s:otherprogs}

\begin{figure} 
    \includegraphics[width=\figw]{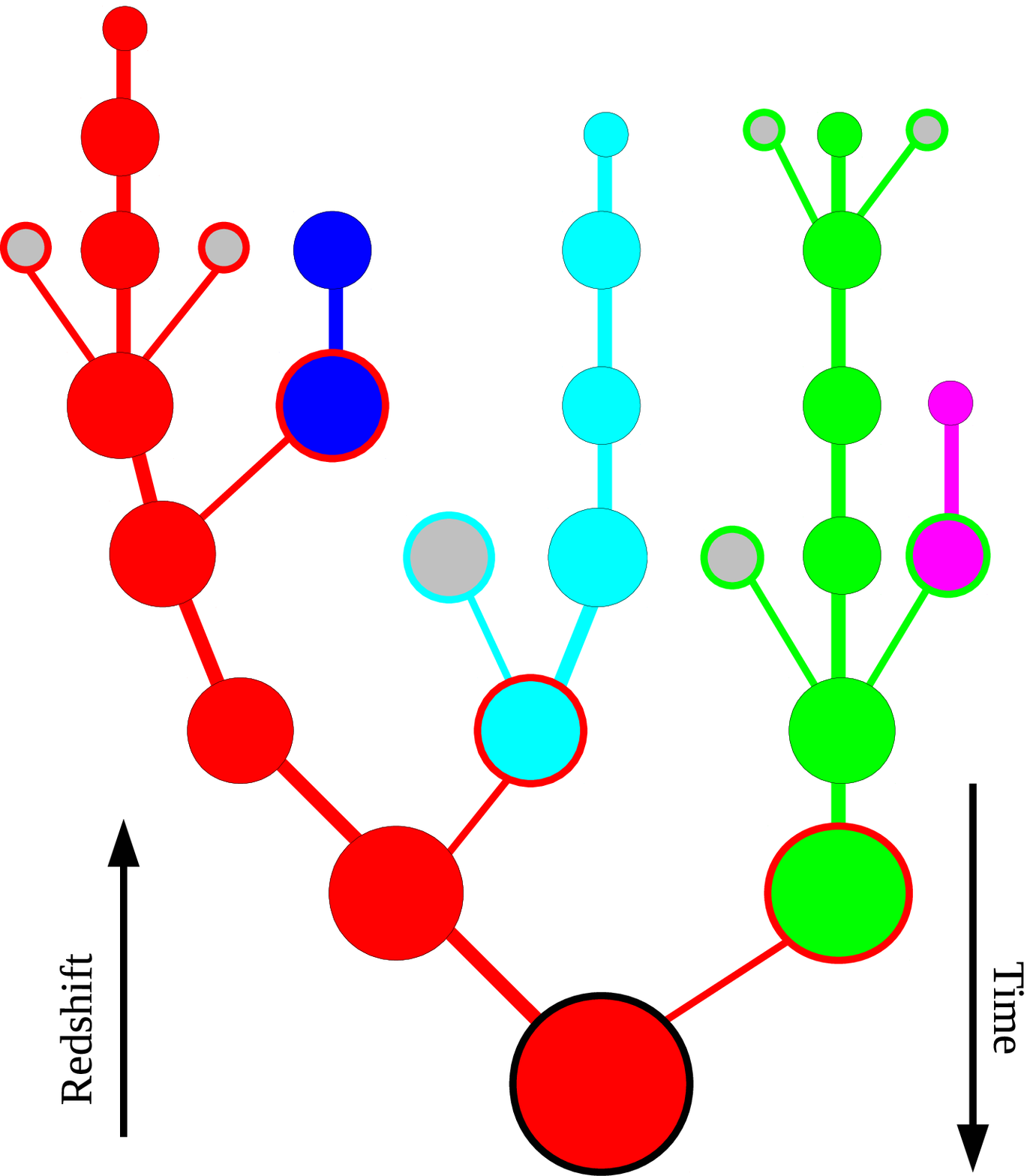}
    \caption{Schematic of a single merger tree, showing immediate
      progenitors of evolutionary tracks (cf. Fig.~\ref{f:mergtree}).
      The circles represent haloes in the tree at different
      timesteps. Each evolutionary track is given an individual colour
      (although `tracks' that only exist for a single timestep are all
      coloured grey).  Haloes at the end-point of their tracks have a
      heavy outline.  When a halo is an immediate progenitor of a halo
      from a different track, it has that track's colour as an
      outline.  
}
  \label{f:mergtreeprogs}
\end{figure}

Halo models are commonly used to study the evolution and statistical
properties of structures, both in the context of galaxy formation and
cosmology in general, such as through Halo Occupation Distribution
models \citep{2000MNRAS.311..793B, 2002ApJ...575..587B} or the
Extended Press--Schechter formalism \citep[e.g.][and references
  therein]{2014MNRAS.440..193J}. In such models, the evolution of halo
properties such as mass is usually assumed to be solely due to mergers
with other haloes.  For example, the mass of a halo at one timestep is
equal to the sum of the masses of its immediate progenitors at the
preceding timestep. Similarly, since angular momentum is a conserved
quantity, one might assume in such models that the halo angular
momentum vector is also equal to the sum of the angular momenta of its
immediate progenitor haloes.  Here, we are able to test the extent to
which this assumption holds for the haloes and merger trees we have
defined.  We illustrate the relationship between immediate progenitors
and halo tracks in Fig.~\ref{f:mergtreeprogs}; in this section we will
be measuring changes in halo properties between a halo at a given
timestep and the sum of the property over its immediate
progenitors. We consider just the root tracks, to avoid
double-counting haloes as end-points of one track and progenitors of
another.  We also use each simulation snapshot output, rather than
interpolating between snapshots to get a constant timescale $\tau$ as
in the previous sections.

\begin{figure*}
  \centering
  \includegraphics[width=\figwthree]{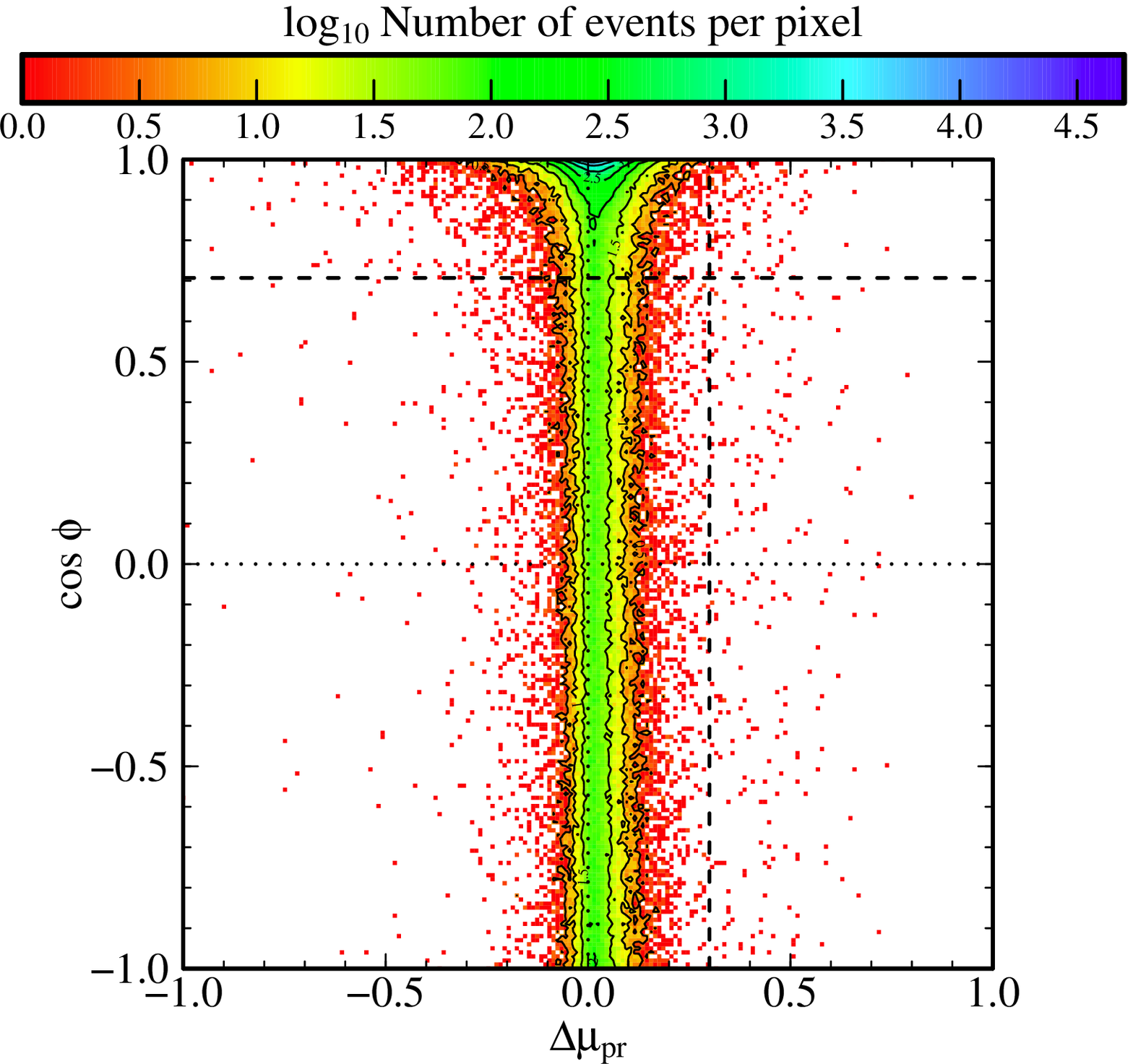}
  \includegraphics[width=\figwthree]{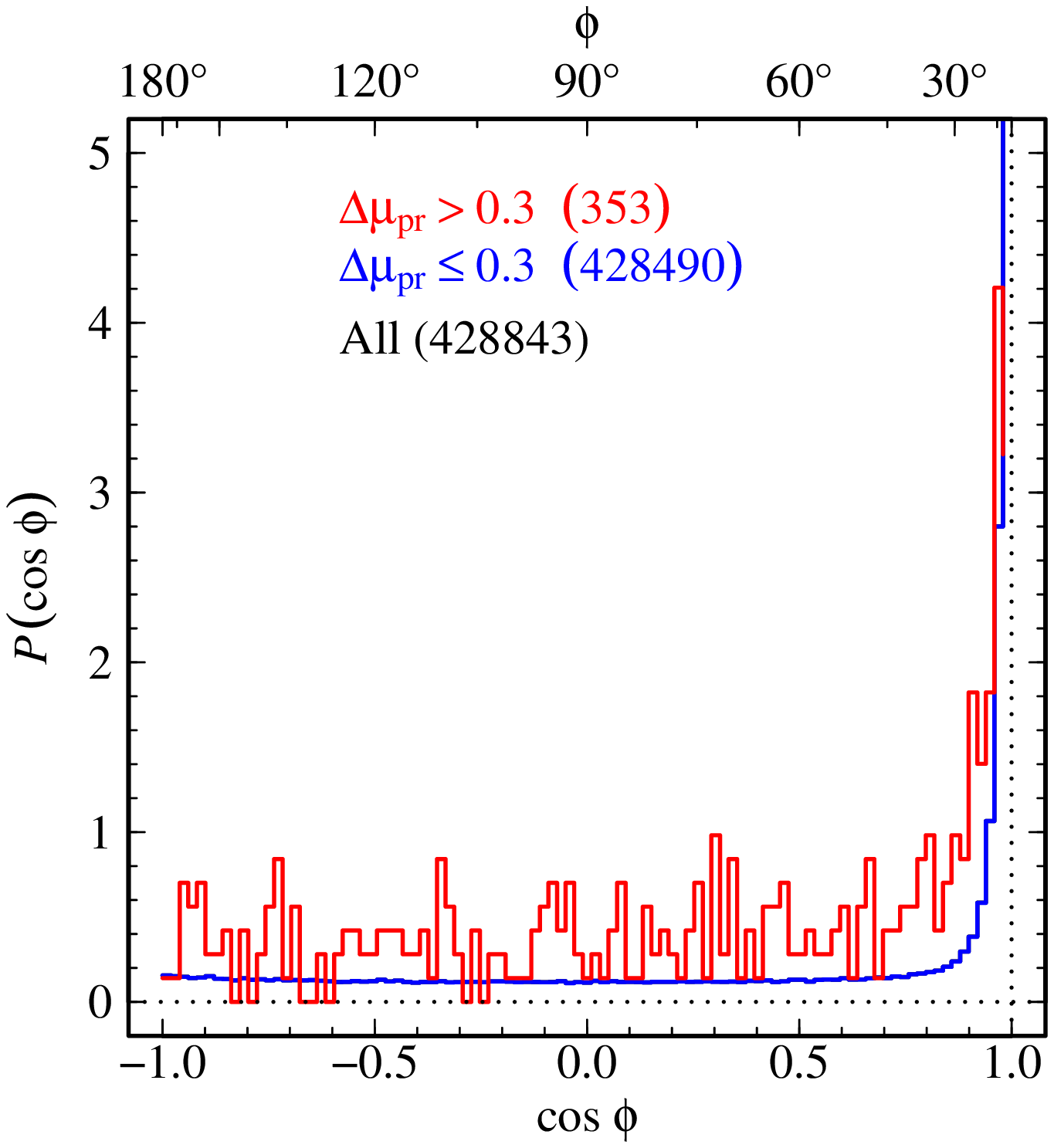}
  \includegraphics[width=\figwthree]{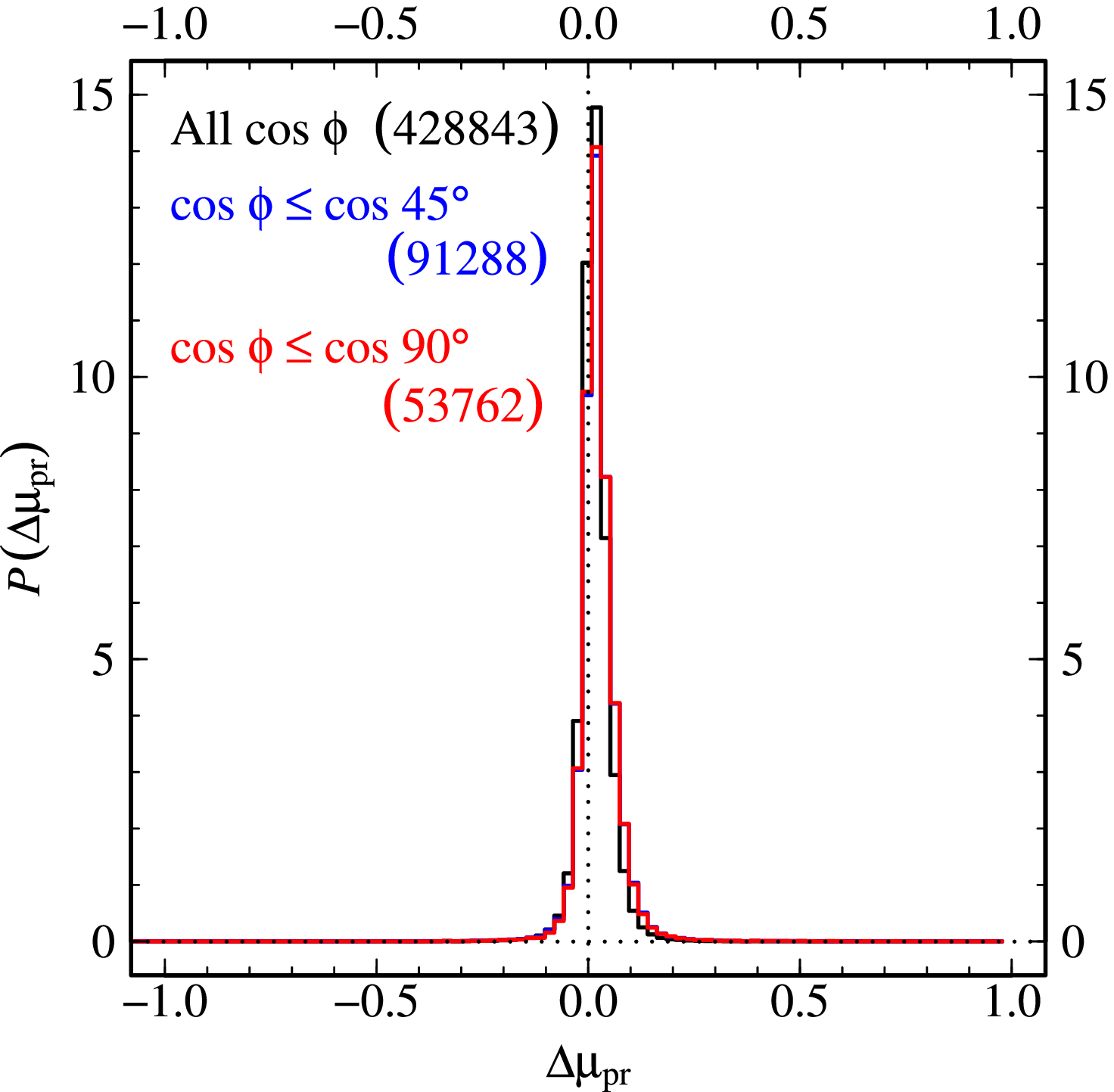}
  \caption{Left: joint distribution of the fractional mass change and
    spin orientation change between haloes at a given timestep and
    their progenitors at the immediately preceding timestep.  Middle:
    histogram showing the distribution across $\cos\phi$, for all
    events (black), and those when there is a large additional change
    in mass (red), and when there is not (blue).  Right: histogram
    showing the cross-section through the joint distribution as a
    function of $\DMpr$, for all events (black), and events when the
    total spin changes by at least $45\degr$ (blue) and $90\degr$
    (red). 
 }
 \label{f:normprogs}
\end{figure*}

We compute the fractional mass change between a halo at time $t_i$
and the sum of the masses of its $\Nprog$ immediate progenitors
identified at the preceding timestep, $t_{i-1}$:
\begin{equation}
  \DMpr(t_i) := \frac{M(t_i) - \sum_{p=1}^{\Nprog}M_p(t_{i-1})}{M(t_i)}.
\end{equation}
We can also compute the centre-of-momentum frame for the set of
progenitors, and thus their total angular momentum vector in that
frame $\vv{J}_\text{progs} = \sum_{p=1}^{\Nprog} \vv{J}_p$.  We
can therefore calculate the change in orientation between that total
angular momentum of the progenitors and the subsequent halo angular
momentum,
\begin{equation}
  \cos\phi(t_i) := \frac{ \vv{J}(t_i)  \dotprod  \vv{J}_\text{progs}(t_{i-1}) }
        {\left| \vv{J}(t) \right|  \left| \vv{J}_\text{progs}(t_{i-1}) \right|}.
\end{equation}
In Fig.~\ref{f:normprogs}, we plot the joint distribution of events in
terms of $\cos\phi$ and $\DMpr$ (subject to the same standard
selection criteria as before, see section \ref{s:sel}), along with
associated histograms (by analogy to our previous figures).

From these figures, we can see that the assumption that mass is
conserved between haloes and their progenitors is quite well-founded,
i.e. $\DMpr \approx 0$.  There are relatively few events with large
values of $|\DMpr|$, with over $84\%$ of the events located within
$|\DMpr|<0.05$.  

The angular momentum orientation behaves differently.  Although it
cannot be seen clearly in the joint distribution image (left panel of
Fig.~\ref{f:normprogs}), the histogram of $\cos\phi$ (middle panel)
shows that there is a very strong peak at no change; we find that
$40\%$ of the events involve orientation changes of less than
$5\degr$.  However, there appears to be an almost uniform probability
for a change in the orientation of the halo spin with respect to its immediate
progenitors, as long as the change is $\phi\ga 45\degr$.
\pbnew{Note that, as mentioned earlier, multiple rapid mergers (between two snapshots) could result in rather chaotic changes in spin direction, as the result depends on the details of the spins and accretion directions of the merging objects. The necessity of using discrete output times means that our results can be seen as a lower limit, and could be quantitatively different with different time resolutions. } 



Thus, while halo mass is reasonably well conserved between timesteps,
angular momentum is not.  This is in agreement with
\cite{2011MNRAS.411.1963B}, who  studied the evolution of angular
momentum of individual haloes and their particles in detail.


\section{Conclusions}\label{s:concs}
In this paper, we have investigated the frequency of changes in the
spin direction of dark matter haloes over their lifetimes, and how
such changes relate to the halo merger history. In a simple halo
model, one might assume that large, sudden changes in spin direction
-- spin flips -- occur exclusively during major mergers, with the
angular momentum direction remaining relatively stable during intervening
times.  Extending the work of \cite{paperI}, we have shown that this
is not the case for haloes with final masses spanning $\sim 10^{11}$
-- $10^{15}\munit$, with spin flips, in fact, occurring often without
major mergers.

We find that $39\%$ of major mergers coincide with spin flips of
$45\degr$ or more. Minor mergers or accretion events are very unlikely
to coincide with such flips: just $2\%$ of non-major-merger events do
so.  However, the shape of the joint distribution of fractional mass
change, $\DM$, and spin orientation change, $\cos\theta$, is such that
large spin flips are very likely to coincide with non-major-merger
events; $76\%$ of spin flips do so.

Changes in spin direction correlate poorly with changes in specific
angular momentum magnitude.  Spin flips coincide with a broad range of
changes in specific angular momentum, albeit with a
slight preference for an increase with large flips.

If we consider only those haloes that survive to $z=0$, then we find
that the joint distribution of $\DM$ and $\cos\theta$ is noticably
narrower in $\DM$: haloes that are doomed to merge into another halo
before $z=0$ have many more mass-loss events ($\DM < 0$). This is probably a
feature of the merger trees we are using, with haloes losing mass
before the timestep at which they cease to be recognised as an
independent halo.  For those haloes that do survive to $z=0$, we find
that less than $1\%$ of minor mergers coincide with large spin
flips, but over $95\%$ of large flips coincide with minor mergers.

Since changes in the inner regions of a halo are more likely to have
a strong impact on the evolution of the central galaxy, we have also
investigated the relationship between flips in the inner halo spin and
mergers in the halo as a whole.  In this case, we find that there is
a general increase in the probability of a spin flip. In particular over
$95\%$ of large inner flips coincide with minor mergers (over $99\%$
for the haloes that survive to $z=0$), and $6.3\%$ of minor mergers
coincide with a large inner flip.

Many of those large flips that coincide with minor mergers do in fact
have a major merger associated with them -- but at a slightly earlier
or later timestep.  We have investigated the number of large flip
events that have a major merger within a given time window $\Delta t$,
and while major mergers can occur before or after large flips, there is
a tendency for major mergers to precede the flip.  However, even
allowing for these inexact conincidences of major mergers and flips,
most large flips nevertheless are \emph{never} associated with a major
merger, even when $\Delta t$ is extended to several gigayears.

As in \citetalias{paperI}, we have also considered the likelihood of a
large spin flip occurring over the lifetime of a halo, in addition to
the distribution of flip events.  We find that $37.5\%$ of
haloes undergo flips of at least $45\degr$ over a timescale of $0.5\Gyr$
($64.1\%$ for inner-halo flips).  Furthermore, despite the broad range
of final masses for the haloes we consider ($\sim
10^{11}$--$10^{15}$), there is little sign of any significant trend in
these results with halo mass.

Finally, we have tested how well\pbnew{-}conserved halo properties are when
going from the set of immediate progenitors at one timestep to the
single resulting halo at the next timestep.  Halo models usually
assume that halo mass is conserved during mergers (i.e. that the sum
of masses of the progenitors equals the mass of the resulting halo),
and we find that this is a reasonably good approximation.  One might
also imagine that the resultant halo's angular momentum is equal to
the (vector) sum of those of its immediate progenitors. However, we do
\emph{not} find this to be the case.  Although $40\%$ of events do
have no orientation change between the net spin of progenitors and the
final halo spin, the distribution in this orientation change becomes
uniform for changes greater than about $45\degr$ -- i.e. all post-merger
orientations greater than $45\degr$ are equally probable.

Our findings have consequences for the use of simple halo models, both
in theoretical studies and when interpreting observations.  Care must
be taken when making modelling assumptions related to angular momentum
in various contexts. These include the relationship between angular
momentum and galaxy morphology; the orientation and persistence of the
orientation of haloes with respect to galaxies, and with respect to
larger-scale structures; and studies that relate dynamical disturbance
solely to galaxy mergers.

The present study has not addressed the cause of the non-major-merger
spin flips that we have observed (although they are presumably related
to flybys of satellite haloes, or similar phenomena), and it would be
interesting to relate them to properties of the immediate environment
of the halo in question.  Although we have treated mergers (and for
the most part, spin flips) as discrete instantaneous events (albeit
with a given timescale $\tau$), with higher time resolution one would
hope to be able to resolve the merger or flip process itself and be
able to measure their timescales directly.

\pbnew{While our choce of algorithms for merger tree, halo definition and selection will have quantitatively affected our results, qualitatively speaking the}
\pbscrap{The} work presented here can be seen as a warning against using
oversimplified halo models of structure and galaxy formation.  Haloes
can clearly be disturbed by processes related to, but separate from,
simply their mass accretion history.  Incorporating additional halo
properties such as spin direction in models of galaxy formation \pbnew{ -- perhaps by tracking spin vectors in addition to halo mass -- }
\pbscrap{(e.g. Padilla et al. 2014)} can prove a useful approach to
improving their ability to match the observed universe.

\pbnew{Due to the difficulties in robustly resolving spin changes in simulations (which motivated our series of selection criteria), \cite{2014MNRAS.443.2801P} took a statistical approach when implementing  spin direction changes in their semi-analytic model, rather than tracking halo spin vectors directly.  Further studies of the statistics of spin flips should be carried out in simulations that resolve  the  dynamics of structures on different scales (at higher resolution, and at larger scales),  to improve and constrain the implementation of spin flips in semi-analytic galaxy formation models.}


\section*{Acknowledgements}
PEB thanks Peter Schneider \& Christiano Porciani for helpful
discussions, and acknowledges the support of the Deutsche
Forschungsgemeinschaft under the project SCHN 342/7--1 in the
framework of the Priority Programme SPP-1177, and the Initiative and
Networking Fund of the Helmholtz Association, contract HA-101
(``Physics at the Terascale''). 
The simulations used in this paper were carried out as part of the
programme of the Virgo Consortium on the Regatta supercomputer of the
Computing Centre of the Max-Planck-Society in Garching, and the
Cosmology Machine supercomputer at the Institute for Computational
Cosmology, Durham.
This work was supported in part by European Research Council (grant
numbers GA 267291 ``Cosmiway'') and by an STFC Consolidated grant to
Durham University. It used the DiRAC Data Centric system at Durham
University, operated by the Institute for Computational Cosmology on
behalf of the STFC DiRAC HPC Facility
(\url{http://www.dirac.ac.uk}). This equipment was funded by BIS
National E-infrastructure capital grant ST/K00042X/1, STFC capital
grant ST/H008519/1, and STFC DiRAC Operations grant ST/K003267/1 and
Durham University. DiRAC is part of the National E-Infrastructure.




\newcommand\repprogphys{Rep.~Prog.~Phys.}  

\bibliographystyle{mnras}
\bibliography{spinflips_both}

\appendix

\section{Choice of halo event timescale}\label{s:timescales}
We wish to find an appropriate characteristic timescale of haloes to use
for calculating changes in halo properties.  While we expect such a
timescale to depend on cosmology, redshift, and halo mass, we aim to
find a \emph{single} value that we can use throughout our analysis.
We will compare our results at different timescales to demonstrate the
degree to which they depend on this choice of timescale.

In general, we can write a timescale as $\tau = \ell/v$.  We take the
characteristic length $\ell$ to be some multiple of a characteristic
halo radius $y r$, where for example $y=2$, $\upi$, $2\upi$, etc, making
$\ell$ the diameter, circumference, etc.  We take the characteristic
velocity $v$ to be the circular velocity of a test particle at $r$
orbiting under gravity, $v=\sqrt{G M(<r)/r}$. The timescale is
therefore given by
\begin{equation}
  \tau = y\sqrt{\frac{r^3}{G M(<r)}}.
\label{e:tscale}
\end{equation}

If we model a halo as the mass within a particular spherical boundary
set by an overdensity criterion, and we take $r=\Rvir$, then $M(<r)
\equiv \Mvir = \frac{4}{3}\upi \Rvir^3 \Deltac(z)\, \rhoc(z)$.  Defining
for convenience $\chi(z) = \OmegaMzero a^{-3} + \OmegaLzero$, we can
write $\rhoc = \chi(z)\, \rhoczero = \chi(z)\, 3H_0^2 / 8\upi G$,
giving
\begin{equation}
  \tau(z) 
  = y\sqrt{\frac{3}{4\upi G} \frac{8\upi G}{3H_0^2} \frac{1}{\chi(z)\Deltac(z)} }
  \,\equiv \frac{y}{H_0}\sqrt{\frac{2}{\chi(z)\Deltac(z)} }.
  \label{e:tauvir} 
\end{equation}
Since our definition of $\Rvir$ depends only on cosmology and time, so
also does $\tau$ -- losing any dependence on halo properties, making
it a cosmological timescale rather than a halo timescale.  In fact,
since the overdensity criterion, $\Deltac(z)$, originates in the
spherical collapse model of halo formation, this timescale is that of
the haloes that are forming at redshift $z$, rather than of the
spectrum of extant haloes at that redshift.

An alternative is to take the characteristic radius somewhere within
$\Rvir$, thus incorporating the halo density profile and retaining
some dependence on halo scale and history.  If we take the
characteristic radius as that enclosing some fraction of the halo
mass, i.e. $\fM = M(<R_f)/\Mvir$, and $r=R_f$, then
equation~(\ref{e:tscale}) becomes:
\begin{eqnarray}
\tau &=& y\sqrt{ \frac{R_f^3}{G \fM \Mvir}}\\
     &=& y\sqrt{ \frac{3}{4\upi G} \frac{8\upi G}{3H_0^2} \frac{R_f^3}{\fM \Rvir^3} \frac{1}{\chi(z)\Deltac(z)}  }\\
     &=& \frac{y}{H_0} \sqrt{ \frac{R_f^3}{\fM \Rvir^3} \frac{2}{\chi(z)\Deltac(z)} }
     \label{e:taufrac}
\end{eqnarray}
We now assume an NFW density profile \citep{nfw96,nfw97}, 
\begin{equation}
  \rho(r) = \frac{\rhos}{\frac{r}{\rs}\left(1+\frac{r}{\rs}\right)^2}
\end{equation}
where the halo concentration is related to the scale radius through
$c=\Rvir/\rs$ and the characteristic density is
\begin{equation}
  \rhos = \frac{\Deltac\rhoc}{3}
          \frac{c^3}{\left(\ln\left(1+c\right) - \frac{c}{1+c}\right)}.
\end{equation}
The cumulative mass is  $M(<R) = 4\upi \int^R_0 \rho(r) r\;\dd r$, so here
\begin{equation}
  M(<R) = 4\upi\rhos \int^R_0 
  \frac{r^2}{\frac{r}{\rs}\left(1+\frac{r}{\rs}\right)^2} \dd r.
\end{equation}
If we define $x=r/\Rvir$ so that $cx=r/\rs$, we get
\begin{eqnarray}
  M(<\Rvir) = 4\upi\rhos \frac{\Rvir^3}{c^3} 
                         \left( \ln(1+cx) - \frac{cx}{1+cx} \right)\\
            = \frac{4}{3}\upi\Rvir^3 \Deltac(z)\rhoc(z) 
                 \left( \frac{ \ln(1+cx) - \frac{cx}{1+cx} }
                             { \ln(1+c ) - \frac{c }{1+c } } \right)
\end{eqnarray}
The radius fraction, $x$, for a given concentration can then be found by
solving
\begin{equation}
             \ln(1+cx) - \frac{cx}{1+cx} 
= \fM \left( \ln(1+c ) - \frac{c }{1+c } \right).
\end{equation}
In the case of $\fM=1/2$, \cite{2001MNRAS.321..155L} provided a
fitting formula for $x(c)$ and thus the half-mass radius, $\Rhalfm$.

To link this back to the halo mass and radius, we need a
concentration--mass relation for all redshifts; fitting formulae for
such a relation are provided, for example, in
\cite{2011MNRAS.411..584M}.  So, for a given halo mass and desired
fraction $\fM$, at a given redshift, we can find a concentration and
thus $x=R_f/\Rvir$, and finally $\tau$.

\begin{figure*} 
  \centering\includegraphics[width=185mm]{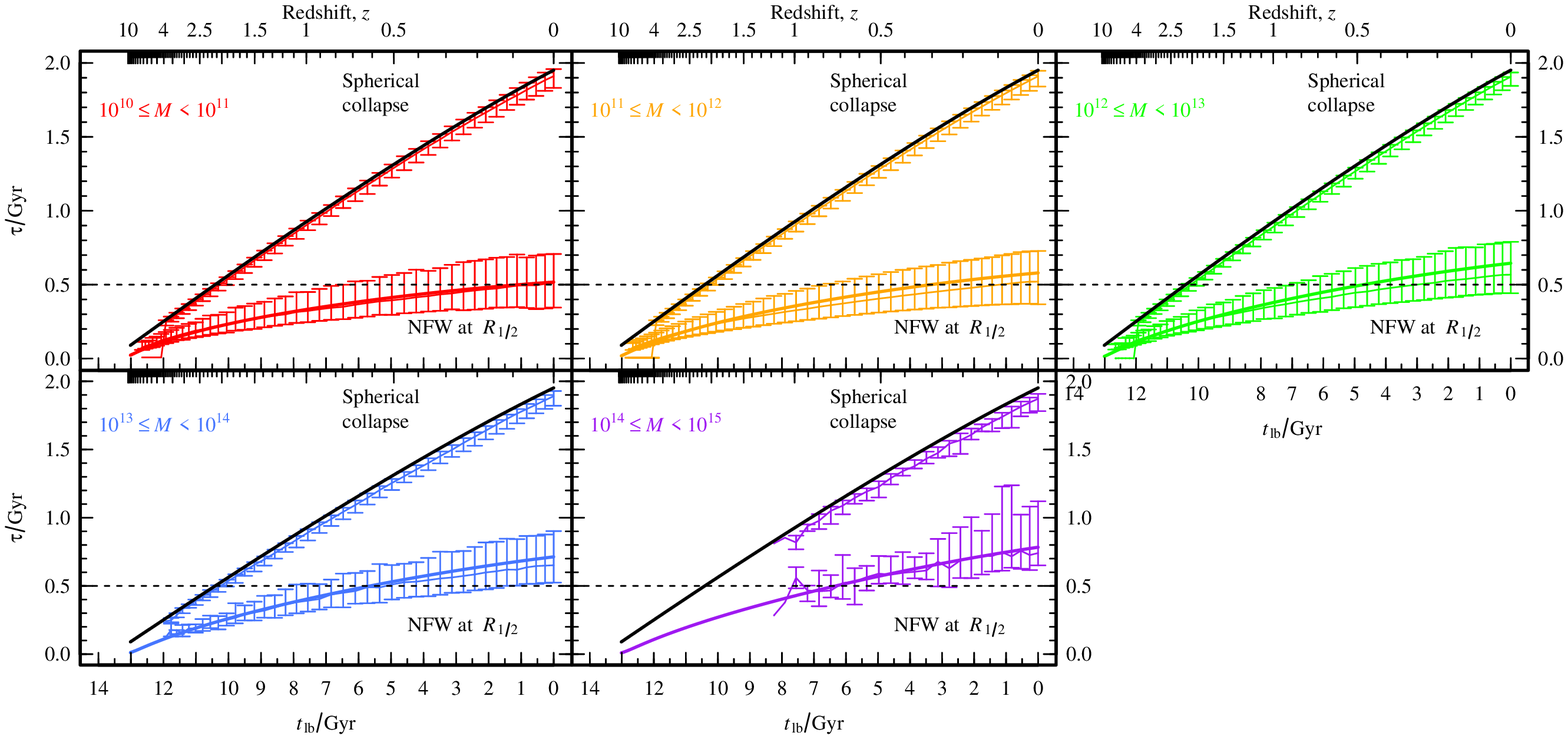}
  \caption{Different halo timescales as a function of lookback time,
    $\tlb$, with each panel showing haloes of different masses at each
    \pbscrap{timestep}\pbnew{snapshot}.  The lines labelled `Spherical collapse' use the
    halo-independent timescale: The analytic result from equation
    (\ref{e:tauvir}) (solid black line) is compared to the medians and
    10th/90th percentiles (points \& error bars) from the distribution
    of measured timescales from the haloes at each \pbscrap{timetep}\pbnew{snapshot}, using
    equation (\ref{e:tscale}) with $r=\Rvir$.  The data labelled `NFW
    at $\Rhalfm$' shows the timescale at the half-mass radius: the
    solid lines are from equation (\ref{e:taufrac}) with $\fM=1/2$,
    assuming an NFW profile and using the fitting formulae (see text;
    \citealt{2011MNRAS.411..584M}, \citealt{2001MNRAS.321..155L}).
    The points again show the medians and percentiles of the measured
    timescale (equation (\ref{e:tscale}) with $r=\Rhalfm$) from haloes
    in mass bins at each \pbscrap{timestep}\pbnew{snapshot}.
 }
  \label{f:timescales}
\end{figure*}

Fig.~\ref{f:timescales} shows the timescales from the above equations
(\ref{e:tauvir}) and (\ref{e:taufrac}) (in the latter taking $\fM=1/2$
so $R_f=\Rhalfm$), along with the timescales \emph{measured} from
haloes at each timestep using $\tau_\text{meas} = \sqrt{\Rvir^3/(G
  \Mh)}$ and $\tau_{1/2,\text{meas}} = \sqrt{2\Rhalfm^3/(G \Mh)}$.
(Haloes are selected in this case only if they have at least $1000$
particles.)  The results from the haloes themselves match the
analytic results very well in all cases.  Given the mass distribution
of our haloes over the time period of interest, we opt to use a single
value of $\tau = 0.5\Gyr$ for analysing  halo property changes.
Using a single value makes for more straightforward comparisons and
analysis as haloes grow and evolve, and the results presented above 
show that this is a reasonable approximation  for our haloes.

\begin{figure*}
  \centering
  \includegraphics[width=\figwthree]{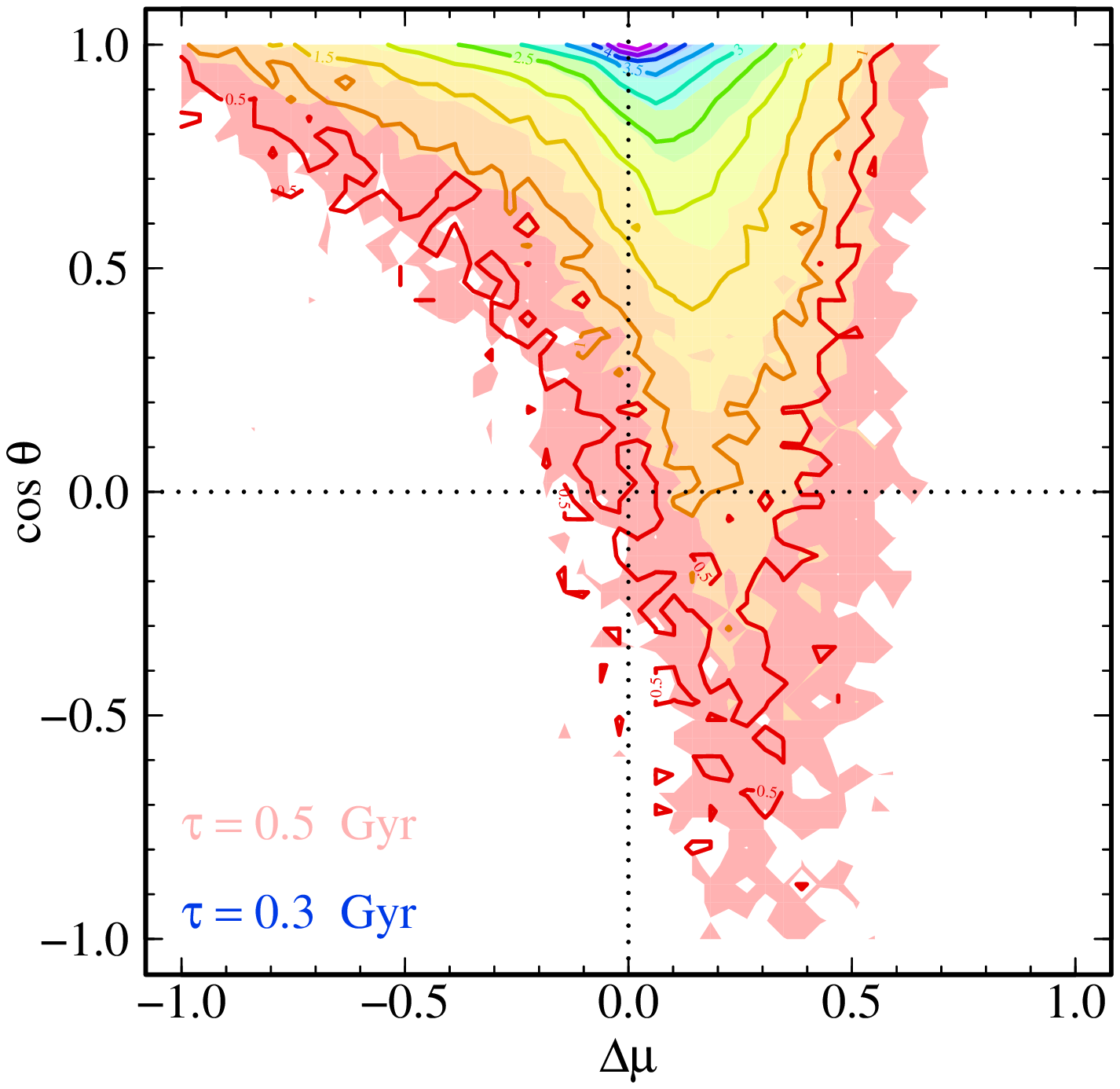}
  \includegraphics[width=\figwthree]{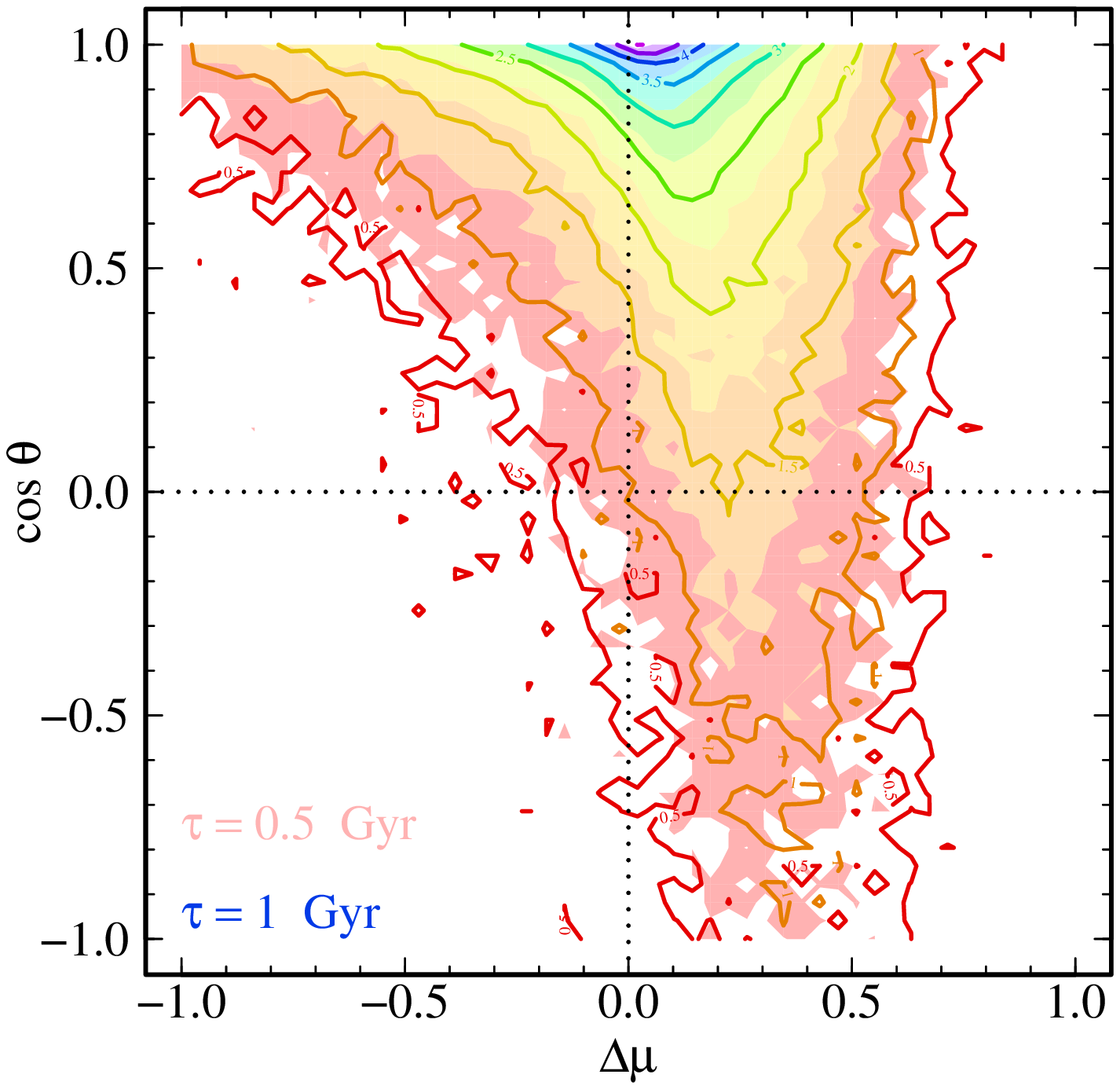}
  \includegraphics[width=\figwthree]{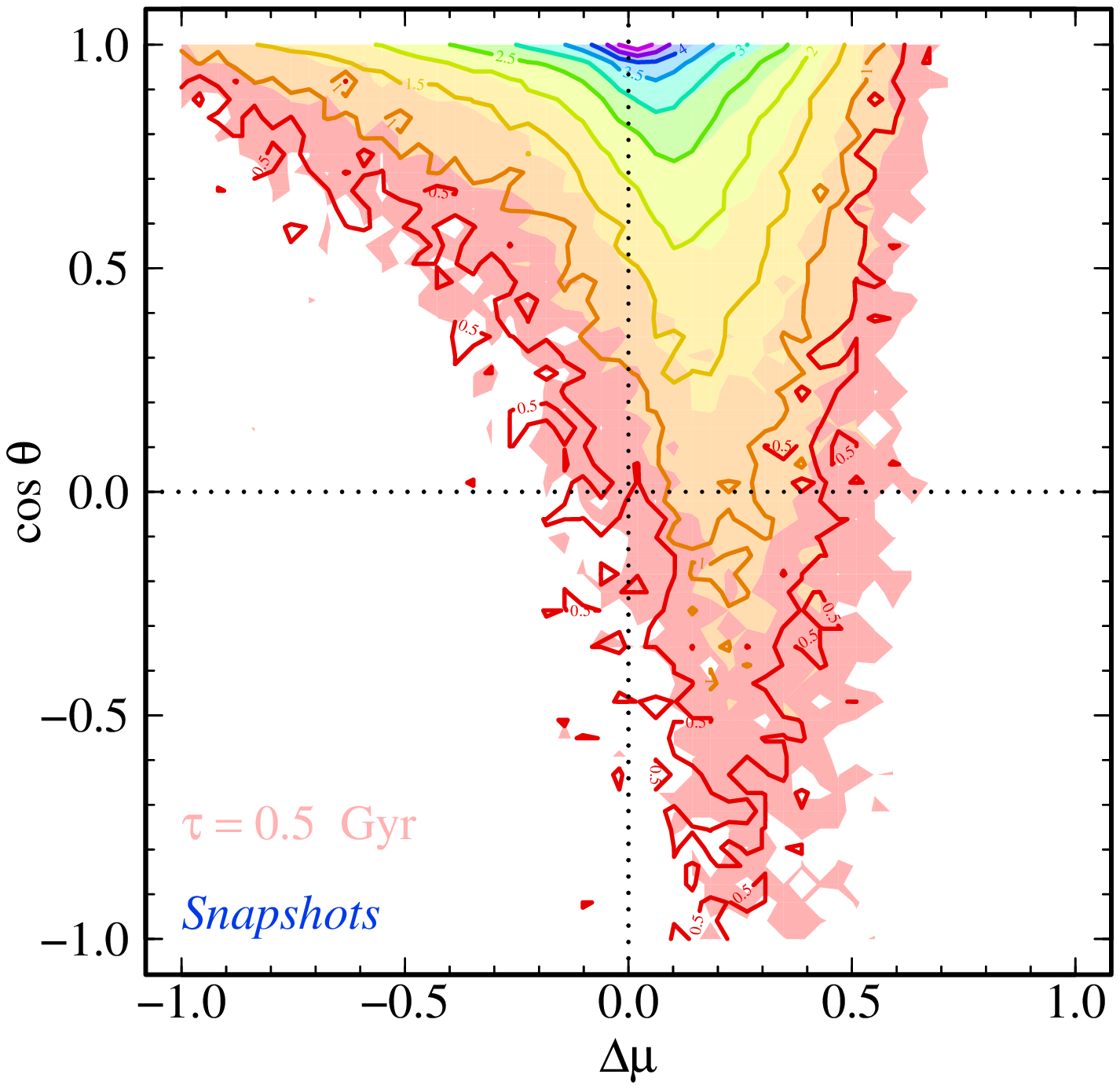}
  \caption{Contour plots of event distributions following that of
    Fig.~\ref{f:distronormalltracks}.  Those results are shown as filled contours.  The outline contours show the results of varying the choice of $\tau$, with the left panel using a shorter timescale, the
    middle panel using a longer timescale. The right panel uses the
    time difference between the simulation snapshots directly, without
    interpolation (i.e. a variable timescale between $~0.4\Gyr$ and
    $0.1\Gyr$.).  The same selection criteria as in
    Fig.~\ref{f:distronormalltracks} are used in all cases.
}
  \label{f:seltestF}
\end{figure*}

We expect the choice of timescale to affect our results
quantitatively, but not qualitatively.  While some of our results are
already shown as a function of measurement timescale (see section
\ref{s:hflipfracDt}), we show here how the event distribution varies
with the choice of $\tau$.  Fig.~\ref{f:seltestF} compares the event
distribution from Fig.~\ref{f:distronormalltracks} (i.e. using $\tau =
0.5\Gyr$), with that obtained when values of $\tau = 0.3\Gyr$ and
$1.0\Gyr$ are used, and when just the (variable) inter-snapshot
timestep is used.  These show that, in practice, there is very little
difference between the universal timescale we choose and just looking
at property differences between snapshots.  In contrast, using a
longer timescale produces a broader distribution, with more events
spread away from the `no change' point, and using a shorter timescale
produces a tighter distribution.  It seems that spin orientation
change $\cos\theta$ is affected more than the fractional mass change
$\DM$.

\bsp	
\label{lastpage}
\end{document}